\documentclass{elsart}
\usepackage{graphics}
\usepackage{epsfig}

\newcommand{\GeV}{\; \mathrm{GeV}}

\newcommand{\lapproxeq}{\lower .7ex\hbox{$\;\stackrel{\textstyle
<}{\sim}\;$}}
\newcommand{\gapproxeq}{\lower .7ex\hbox{$\;\stackrel{\textstyle
>}{\sim}\;$}}
\newcommand{\stackdown}[2]{\lower 1.4ex\hbox{$\;\stackrel{\textstyle{#1}}
{\scriptstyle{#2}}\;$}}
\newcommand{\beq}{\begin{equation}}
\newcommand{\eeq}{\end{equation}}
\newcommand{\bea}{\begin{eqnarray}}
\newcommand{\eea}{\end{eqnarray}}

\def\beq{\begin{equation}}
\def\eeq{\end{equation}}

\makeatletter
\def\slash{\@ifnextchar[{\fmsl@sh}{\fmsl@sh[0mu]}}
\def\fmsl@sh[#1]#2{  \mathchoice
    {\@fmsl@sh\displaystyle{#1}{#2}}    {\@fmsl@sh\textstyle{#1}{#2}}
{\@fmsl@sh\scriptstyle{#1}{#2}}    {\@fmsl@sh\scriptscriptstyle{#1}{#2}}}
\def\@fmsl@sh#1#2#3{\m@th\ooalign{$\hfil#1\mkern#2/\hfil$\crcr$#1#3$}}
\makeatother

\def\beq{\begin{equation}}
\def\eeq{\end{equation}}

\catcode`\@=11

\def\lsim{\mathrel{\mathpalette\@versim<}}
\def\gsim{\mathrel{\mathpalette\@versim>}}
\def\@versim#1#2{\vcenter{\offinterlineskip
    \ialign{$\m@th#1\hfil##\hfil$\crcr#2\crcr\sim\crcr } }}

\def\t1{{\tilde 1}}

\def\slash#1{#1\hskip-6pt/\hskip6pt}

\def\GeV{\,{\rm GeV}}

\def\to{\rightarrow}

\def\sv{\left<\sigma v\right>}

\begin{document}

\begin{frontmatter}



\title{May Heavy hadrons of the 4th generation \\ be hidden in our Universe while close to detection?}
\author{K.Belotsky}
\address{Centre for CosmoParticle Physics "Cosmion", 125047
Moscow, Russia\\  Moscow Engineering Physics Institute, 115409
Moscow, Russia}
\ead{k-belotsky@yandex.ru}

\author{D.~Fargion $*$}
\address{Physics department, Universita' degli studi "La Sapienza", \\
         Piazzale Aldo Moro 5,CAP 00185 Roma, Italy and \\
        INFN Roma, Istituto Nazionale di Fisica Nucleare, Italy}
\corauth[cor]{Corresponding author.}
\ead{daniele.fargion@roma1.infn.it}

\author{M.Yu.~Khlopov}
\address{Centre for CosmoParticle Physics "Cosmion", \\
Keldysh Institute of Applied Mathematics, 125047,
Moscow, Russia\\  Moscow Engineering Physics Institute, 115409
Moscow, Russia \\ Physics Department, University Rome~1 La Sapienza\\
Ple A. Moro 2, 00185 Rome, Italy}
\ead{Maxim.Khlopov@roma1.infn.it}

\author{R.V.~Konoplich}
\address{Department of Physics, New York University, \\
New York, NY 10003, USA \\
Department of Physics, Manhattan College, Riverdale, New
York, NY 10471, USA \\Physics Department, University Rome~1 La Sapienza\\
Ple A. Moro 2, 00185 Rome, Italy}
\ead{rk60@nyu.edu}

\author{M.G. Ryskin}
\address{Petersburg Nuclear Physics Institute, Gatchina,
St.~Petersburg, 188300, Russia}
\ead{ryskin@MR11084.spb.edu}

\author{K.I.Shibaev}
\address{Centre for CosmoParticle Physics "Cosmion", 125047
Moscow, Russia\\  Moscow Engineering Physics Institute, 115409
Moscow, Russia}
\ead{shibaev01@yandex.ru}

\begin{abstract}
Metastable quarks of 4th generation are predicted in the
framework of heterotic string phenomenology. Their presence in
heavy stable hadrons are usually strongly constrained;  however their
$hidden$ compositions in Heavy doubly charged baryons here
considered  are  found to be still allowable: we studied
 their primordial quark production in the early Universe, their freezing into
cosmic Heavy hadrons, their later annihilation into cosmic ray as
well as their relic presence in our Universe and among us on
Earth. We discuss also their possible production in present or
future accelerators. Indeed if the lightest quarks and antiquarks
of the 4th generation are stored in doubly charged baryons and
neutral mesons, their lifetime can exceed the age of the Universe;
the existence of such an anomalous Helium-like (and neutral
Pion-like) stable particles may escape present experimental
limits, while being close to present and future experimental
test. On the contrary primordial abundance of lightest hadrons of
the 4th generation with charge $+1$ can not decrease below the
experimental upper limits on anomalous hydrogen and therefore (if
stable) it is excluded. While 4th quark hadrons are rare, their
presence may play different and surprising role in cosmic rays,
muon and neutrino fluxes and cosmic electromagnetic spectra. Most
of these traces are tiny, but nearly detectable.
\end{abstract}

\begin{keyword}
cosmology \sep cosmic rays \sep 4th generation \sep anomalous helium

\PACS 98.80.Cq \sep 04.70.Bw \sep 04.70.Dy \sep 98.80.Hw(?)

\end{keyword}

\end{frontmatter}

\section{\label{introduction}Introduction}

The problem of existence of quarks and leptons of 4th generation
is among the most important in the modern high energy physics.
The natural extension of the
Standard model leads in the heterotic string
phenomenology to the prediction of fourth generation of quarks and
leptons \cite{Shibaev,Sakhenhance} with a stable massive 4th neutrino
\cite{Fargion99,Grossi,Belotsky,BKS}. The comparison between the rank of the
unifying group $E_{6}$ ($r=6$) and the rank of the Standard model
($r=4$) implies the existence of new conserved charges and new
(possibly strict) gauge symmetries. New strict gauge U(1) symmetry
(similar to U(1) symmetry of electrodynamics) is possible, if it
is ascribed to the fermions of 4th generation. This hypothesis
explains the difference between the three known types of neutrinos
and neutrino of 4th generation. The latter possesses new gauge
charge and, being Dirac particle, can not have small Majorana mass
due to see saw mechanism. If the 4th neutrino is the lightest
particle of the 4th quark-lepton family, strict conservation of
the new charge makes massive 4th neutrino to be absolutely stable.
Following this hypothesis \cite{Shibaev} quarks and leptons of 4th
generation are the source of new long range interaction,
(which we'll call further $y$-interaction) similar to the electromagnetic interaction
of ordinary charged particles.

Recent analysis \cite{Okun} of precision data on the Standard
model parameters has taken into account possible virtual
contributions of 4th generation particles. It was shown that 4th
quark-lepton generation is not excluded if 4th neutrino, being
Dirac and (quasi-)stable, has a mass about 50 GeV (47-50 GeV is
$1\sigma$ interval, 46.3-75 GeV is $2\sigma$ interval)
\cite{Okun} and other 4th generation fermions satisfy their
direct experimental constrains (above 80-220 GeV). We will
concentrate our attention here on given values of mass for 4th
generation fermions, assuming that the mass of the lightest quark
is about 250 GeV.

If the lightest quark $Q$ of 4th generation possess
new strictly conserved gauge charge, it can decay only
to 4th generation leptons owing to GUT-type interactions,
what makes it sufficiently long living. If its lifetime exceeds the age of the Universe,
primordial $Q$-quark (and $\bar Q$-quark) hadrons should be present in the modern matter.
If this lifetime is less than the age of the Universe, there should be no
primordial 4th generation quarks, but they can be produced
in cosmic ray interactions and be present in cosmic ray fluxes.
The assumed masses for these quarks
make their search a challenge for the present and future accelerators.

In principle, the composition of the lightest baryon of 4th generation
can be: $(Uuu)$ (charge $+2$)\footnote{One can call this particle $\Delta^{++}_{U}$.},
$(Uud)$ (charge $+1$), $(Udd)$ (charge $0$) or  (if $m_D<m_U$)
$(Duu)$ (charge $+1$),$(Dud)$ (charge $0$), $(Ddd)$  (charge $-1$)
and the corresponding lightest mesons $(\bar U u)$ (charge $0$), $(\bar U d)$  (charge $-1$)
or $(\bar D u)$ (charge $+1$) and $(\bar D d)$ (charge $0$).
In the present paper we mainly follow the assumption that $m_D>m_U$,
which excludes the baryon with the negative charge, while the lightest
$(\bar U d)$ meson seems to be excluded by the quark model arguments.
These arguments also exclude neutral $(Udd)$ baryon as the lightest.
It leaves theoretically favorable $(Uud)$ and theoretically less favorable $(Uuu)$
as only candidates for lightest baryon of 4th generation
with the only possibility of neutral $(\bar U u)$ as the lightest meson.

We analyze the mechanisms of production of
metastable $Q$ (and $\bar Q$) hadrons in the early Universe, cosmic
rays and accelerators and reveal the possible signatures of their
existence. We'll show that though it may be more probable
that the lightest hadron of the 4th generation has the charge $+1$
and  $Q$-quark should have decay lifetime less than
the age of the Universe to avoid the contradiction with the negative results of the
anomalous hydrogen search, the theoretically less favorable
case of lightest $(Uuu)$ baryon and $(\bar U u)$ meson,
mostly considered in the present paper, deserves special interest.
Such doubly charged and neutral primordial hadrons can be sufficiently stable to be
present in the modern Universe and still remain elusive for their direct searches.
On the other hand the set of direct and indirect effects of their existence
provides the test in cosmic ray and underground experiments
which can be decisive for this hypothesis.
In view of the interesting possibility revealed for $Q$-quark to be sufficiently stable
we mainly use throughout the paper the examples, corresponding to the case,
when $U$-quark of 4th generation is lighter than $D$-quark. However the calculations
of primordial abundance, cosmic ray production and effects of heavy quark annihilation
bear general character and are valid for any charge assignment for $Q$ and its lightest
hadrons.

In the next sections we consider $U$ and $\bar U$ hadrons presence
since early Universe (Sec.2) up to their early freezing
(details in App. 1-4) and later galactic clustering,
their annihilation and survival (Sec.3, Appendix 5), their blowing wind onto Earth (Sec.4)
and their further annihilation and relic presence in cosmic rays (sec.5),
polluted matter, rarest upgoing muons and possible Heavy quark UHECR showers influence (App.5-8).
We consider (Sec.6) possible Heavy quark signatures in Tevatron and LHC.
In the final Section 7 we summarize our present study on the novel
cosmic and astrophysical scenario on Heavy hadrons, possible hidden and
stored just here among us.

\section{\label{primordial} Primordial $U$-hadrons from Big Bang Universe}


In the early Universe at temperatures highly above their masses
fermions of 4th generation were in thermodynamics equilibrium
with relativistic plasma. Strict conservation of new gauge U(1)
charge ($y$-charge) implies $y$-charge symmetry of their
distribution. When in the course of expansion the temperature $T$
falls down below the mass of the lightest $U$-quark, $m$,
equilibrium concentration of quark-antiquark pairs of 4th
generation is given by \beq n_4 = g_4
\left(\frac{Tm}{2\pi}\right)^{3/2} \exp{(-m/T)}, \label{equil}
\eeq where $g_4=6$ is the effective number of their spin degrees
of freedom. We use the units $\hbar = c = k =1$ throughout this
paper.

At the freezing out temperature $T_f$ the rate of expansion
exceeds the rate of annihilation to gluons $U \bar U \rightarrow gg$ and $U \bar U$ pairs are frozen out.
The frozen out concentration at $T \sim T_f \approx m/30$ (See Appendix 1) is given by
\beq
r_4 \approx 2.5 \cdot 10^{-14}\frac{m}{250{\GeV}}.
\label{freez2m}
\eeq
Even this value of primordial concentration of $U$-quarks with the mass $m = 250 GeV$ would lead to the contribution into the modern density,
which is by an order
of magnitude less, than the baryonic density, so that $U$-quarks can not play a significant dynamical role
in the modern Universe. The actual value of primordial $U$-particle concentration should be much smaller
due to radiative and hadronic recombination, which reduce the abundance of frozen out $U$-particles.

Radiation of mutually attracting $y$-charges of $U$ and $\bar U$ can lead to formation of
$U \bar U$ systems bound by $y$-attraction, in which $U$ and $\bar U$ rapidly annihilate.
The rate of this radiative binding is calculated in Appendix 2 and is given by
\beq
\sv \approx 6 \cdot 10^{-13} (\frac{\alpha}{1/30})^{2}(\frac{300K}{T})^{9/10}(\frac{250GeV}{m})^{11/10} \frac{cm^3}{s}.
\label{sigimpact}
\eeq

The decrease of $U$-hadron abundance owing to $U \bar U$ recombination
is governed by the equation
\beq
\frac{dr_4}{dt} = - r_4^2 \cdot s \cdot \sv,
\label{hadrecomb}
\eeq
where $s$ is the entropy density (see Appendix 1).

Its solution at $T_f > T$ (see Appendix 2) turns to be independent on the frozen out concentration (\ref{freez2m})
and is given by (see Appendix 2)
\beq
r_4 \approx 0.013\, \left(\frac{m}{T_f}\right)^{1/10} \frac{m}{\alpha_y^2 m_{Pl}}\approx 4\cdot 10^{-16}
\frac{m}{250\GeV}\left(\frac{30^{-1}}{\alpha_y}\right)^2,
\label{radrecsolq}
\eeq
where $\alpha_y \sim 1/30$ is the running constant of $y$-interaction.

After QCD phase transition at $T = T_{QCD} \approx 150$MeV quarks of 4th generation
combine with light quarks into $U$-hadrons.
In baryon asymmetrical Universe only excessive valence quarks should enter such hadrons,
so that $U$-baryons and $\bar U$-mesons are formed. The details of $U$- and $\bar U$-quark
hadronization are discussed in Appendix 3 and following this discussion we consider
further mainly doubly charged $(Uuu)$ baryon and neutral $(\bar U u)$ meson\footnote{The total
charge of $(Uuu)$ and $(\bar U u)$ is $+2$, what looks like a net electric charge asymmetry
of the Universe. However the net charge of the Universe is vanishing due to the presence of
two corresponding excessive electrons. This pair of electrons compensates the positive charge
of three light $u$ quarks, and the mechanism of this compensation is provided by the usual generation
of baryon (and lepton) asymmetry of the Universe.}.

As it was revealed in \cite{Shibaev,Sakhenhance}
in the collisions of such mesons and baryons recombination
of $U$ and $\bar U$ into unstable $(U \bar U)$ "charmonium -like" state
can take place, thus successively reducing the $U$-hadron abundance.
Hadronic recombination should take place even in the absence
of long range $y$-interaction of $U$-particles. So, we give first
the result without the account of radiative recombination
induced by this interaction.

The uncertainties in the estimation of hadronic recombination rate are discussed in Appendix 4.
The maximal estimation for the reaction rate of recombination $\sv$ is given by
\beq
\sv \sim \frac{1}{m_{\pi}^2} \approx 6 \cdot 10^{-16}\, {\rm \frac{cm^3}{s}}
\label{hadsigmv}
\eeq
or by
\beq
\sv \sim \frac{1}{m_{\rho}^2} \approx 2 \cdot 10^{-17} \frac{cm^3}{s}.
\label{hadsigmv1}
\eeq
The minimal realistic estimation (See Appendix 4) gives
\beq
\sv \approx 0.4 \cdot (T_{eff} m^3)^{-1/2} (3 + \log{(T_{QCD}/T_{eff})}),
\label{hadrecmin}
\eeq
where $T_{eff} = \max{\{T, \alpha_y m_{\pi}\}}$.

Solution of Eq.(\ref{hadrecomb}) for $\sv$ from the Eq.(\ref{hadsigmv}) is given by

Case A
\beq
r_4 = \frac{r_0}{1 + r_0\cdot \sqrt{\frac{\pi g_{QCD}}{45}} \frac{m_{Pl}}{m_{\pi}}
\frac{T_{QCD}}{m_{\pi}} } \approx 1.0\cdot 10^{-20}
\label{hadrecsol}
\eeq
and it is $(\frac{m_{\rho}}{m_{\pi}})^2 \sim 30$ times larger for $\sv$ from
the Eq.(\ref{hadsigmv1}):

Case B
\beq
r_4 = \frac{r_0}{1 + r_0\cdot \sqrt{\frac{\pi g_{QCD}}{45}} \frac{m_{Pl}}{m_{\rho}}
\frac{T_{QCD}}{m_{\rho}} } \approx 3.0 \cdot 10^{-19}.
\label{hadrecsol2}
\eeq
For the minimal estimation of recombination rate (\ref{hadrecmin})
the solution of Eq.(\ref{hadrecomb}) has the form

\beq
r_4 = \frac{r_0}{1 + r_0 2 \cdot \sqrt{\frac{\pi g_{QCD}}{45}} \frac{m_{Pl}}{m}
\sqrt{\frac{T_{QCD}}{m}} } \approx 5 \cdot 10^{-16}(\frac{m}{250 GeV})^{3/2},
\label{hadrecsol3}
\eeq
where in all the cases $r_0$ is given by Eq.(\ref{freez2m}) and $g_{QCD}\approx 15$ (see Appendix 3).
These solutions are independent on the actual initial value of $r_4 = r_0$, if before
QCD phase transition it was of the order
of (\ref{freez2m}). We neglect in our estimations possible effects of recombination in the intermediate period,
when QCD phase transition proceeds.

With the account for radiative recombination it is the value (\ref{radrecsolq}) that should be taken
at $T \sim T_{QCD}$ as $r_0$ in the solutions (\ref{hadrecsol}), (\ref{hadrecsol2}) and
(\ref{hadrecsol3}) for the results of hadronic recombination. Such account does not change the values
(\ref{hadrecsol}) and (\ref{hadrecsol2}), which are still with the high accuracy independent on $r_0$.
However, for the minimal estimation of the recombination rate (\ref{hadrecmin}) the result
of hadronic recombination is modified and reads

Case C
\beq
r_4 \approx 2 \cdot 10^{-16}(\frac{m}{250 GeV})^{3/2}.
\label{hadrecsol4}
\eeq


The existence of new massless U(1) gauge boson ($y$-photon) implies the presence
of primordial thermal $y$-photon background in the Universe. Such background should
be in equilibrium with ordinary plasma and radiation until the lightest particle
bearing $y$-charge (4th neutrino) freezes out. For the accepted value of 4th neutrino mass
($\ge 50$GeV) 4th neutrino freezing out and correspondingly decoupling of $y$-photons
takes place before the QCD phase transition, when the total number of effective degrees
of freedom is sufficiently large to suppress the effects of $y$-photon background
in the period of Big Bang nucleosynthesis. This background does not interact with nucleons
and does not influence the BBN reactions rate, while the suppression of $y$-photon energy density
leads to insignificant effect in the speeding up cosmological expansion rate in the BBN period.
In the framework of the present consideration
the existence of primordial $y$-photons does not play any significant role
in the successive evolution of $U$-hadrons.

In the period of recombination of nuclei with electrons charged $U$-baryons
recombine with electrons to form
atoms of anomalous isotopes. The substantial (up to 10 orders
of magnitude) excess of electron number density over the number density
of primordial $U$-baryons makes virtually all $U$-baryons to form atoms.
The cosmological abundance of free charged $U$-baryons is to be exponentially
small after recombination. If the lightest is $(Uuu)$ baryon with
electric charge $+2$, atoms of anomalous He are formed.
$U$-hadrons with charge $+1$ form atoms of anomalous hydrogen.

These atoms, having atomic cross sections
of interaction with matter, participate then in formation
of astrophysical bodies, when galaxies are formed. One might assume that neutral $U$-hadrons,
having nuclear interactions and being clustered in galaxies,
should not follow matter in formation of stars and planets,
and that one can expect suppression of their concentration in such bodies.
However the existence of Coulomb-like $y$-attraction will make them
to obey the condition of neutrality in respect to the $y$-charge.
Therefore owing to neutrality condition the number densities of
$U$- and $\bar U$-hadrons in astrophysical bodies should be equal\footnote{In this sense
there is a very remarkable and obvious difference between common (light-quark) baryon matter
and the 4th quark matter: there is no room for any lepton-baryon asymmetry as in our matter.
The same asymmetry leads to 250 times larger cosmological density for this matter, than for ordinary
baryonic, what is by 10 times larger, than it is allowed for dark matter. The contradiction
grows to 20 orders of magnitude, if we take into account the upper bounds on the presence
of this matter around us.}.

\section{\label{matter} Evolution of $U$-hadron content in galactic matter}
In the astrophysical body with atomic number density $n$ the initial $U$-hadron abundance
$n_{U0} = r_u \cdot n$ can decrease with time due to $U \bar U$ recombination.
Under the neutrality condition
$$n_U = n_{\bar U}$$
the relative $U$-hadron abundance $r = n_U/n= n_{\bar U}/n$ is governed by the equation
\beq
\frac{dr}{dt} = - r^2 \cdot n \cdot \sv.
\label{recomb}
\eeq
The solution of this equation is given by
\beq
r = \frac{r_u}{1 + r_u \cdot n \cdot \sv \cdot t}.
\label{sol}
\eeq

If
\beq
n \cdot \sv \cdot t \gg \frac{1}{r_u},
\label{cond}
\eeq
the solution (\ref{sol}) takes the form
\beq
r = \frac{1}{n \cdot \sv \cdot t}.
\label{sol2}
\eeq
and, being independent on the initial value, $U$-hadron abundance
decreases inversely proportional to time.

By definition $r_u = f_i/A_{atom}$, where $A_{atom}$ is the averaged atomic weight of the considered matter
and $f_i$ is the initial $U$-hadron to baryon ratio. In the pregalactic matter this ratio is determined by $r_4$ from A) Eq.(\ref{hadrecsol}),
B) Eq.(\ref{hadrecsol2}) and C) Eq.(\ref{hadrecsol4}) and is
equal to
\beq
f_4 = \frac{r_4}{r_b} =\left\{
 \begin{array}{c}
 10^{-10} \, {\rm for \, the \, case \, A},\\
 3 \cdot 10^{-9} \, {\rm for \, the \, case \,B},\\
2 \cdot  10^{-6} \, {\rm for \, the \, case \,C}.
 \end{array}
 \right.
\label{primcr1}
\eeq
Here $r_b \approx 10^{-10}$ is baryon to entropy ratio, defined in Eq.(\ref{bard}) in Appendix 3.

Taking for averaged atomic number density in the Earth $n \approx 10^{23}
cm^{-3}$, one finds that during the age of the Solar system
primordial $U$-hadron abundance in the terrestrial matter should
have been reduced down to $r \approx 10^{-28}$. One should
expect similar reduction of $U$-hadron concentration in Sun and
all the other old sufficiently dense astrophysical bodies.
Therefore in our own body we might contain just one of such heavy hadrons.
However, as shown later on in Section 4, the persistent pollution
from the galactic gas nevertheless may increase this relic number density
to a stationary much larger ($r \approx 10^{-23}$) value.

The principal possibility of strong reduction in dense bodies for
primordial abundance of exotic charge symmetric particles
due to their recombination in unstable charmonium like systems
was first revealed in \cite{fractons} for fractionally charged
colorless composite particles (fractons).

The $U$-hadron abundance in the interstellar gas strongly depends
on the matter evolution in Galaxy, which is still not known to the extent,
we need for our discussion.

Indeed, in the opposite case of low density or of short time interval,
when the condition (\ref{cond}) is not valid, namely, at
\beq
n < \frac{1}{r_u \sv t} = A_{atom} \cdot \frac{T}{300K} \cdot \frac{t_U}{t} cm^{-3} \left\{
 \begin{array}{c}
 4 \cdot 10^{4} \, {\rm for \, the \, case \, A},\\
 10^{3} \, {\rm for \, the \, case \, B},\\
2 \, {\rm for \, the \, case \, C},
 \end{array}
\right.
\label{lowcond}
\eeq
where $t_U=4 \cdot 10^{17} s$ is the age of the Universe,
 $U$-hadron abundance does not change its initial value.

In principle,
if in the course of evolution matter in the forming Galaxy
was present during sufficiently long period ($t \sim 10^9 yrs$)
within cold ($T \sim 10K$) clouds with density $n \sim 10^3 cm^{-3}$
$U$-hadron abundance retains its primordial value for the cases A and B ($f_i = f_4$),
but falls down $f_i = 5 \cdot 10^{-9}$ in the case C,
making this case close to the case B.
The above argument may not imply all the $U$-hadrons to be initially present
in cold clouds. They can pass through cold clouds and decrease their abundance
in the case C at the stage of thermal instability,
when cooling gas clouds, before they become gravitationally bound,
are bound by the external pressure of the hot gas. Owing to their large inertia
heavy $U$-hadron atoms from the hot gas can penetrate  much deeply inside
the cloud and can be captured by it much more effectively, than ordinary atoms. Such mechanism can
provide additional support for reduction of
$U$-hadron abundance in the case C.

However, according to Appendix 7,in particular annihilation of $U$-hadrons leads to multiple $\gamma$ production.
If $U$-hadrons with relative abundance $f_4$ annihilate at the redshift $z$, it should leave
in the modern Universe a background $\gamma$ flux
$$F(E>E_{\gamma}) = \frac{N_{\gamma} \cdot f_4 \cdot r_b \cdot n_{\gamma\,mod} \cdot c}{4 \pi}
\approx 3 \cdot 10^3 f_4 (cm^2 \cdot s \cdot ster)^{-1},$$
of $\gamma$ quanta with energies $E>E_{\gamma} = 10 GeV/(1+z)$. Here $ n_{\gamma\,mod} \approx 400 cm^{-3}$
is the modern number density of relic photons and the numerical values for $\gamma$ multiplicity $N_{\gamma}$
are given in table 1 of Appendix 7. So annihilation even as early as at $z \sim 9$
leads in the case C to the contribution into diffuse extragalactic gamma emission,
exceeding the flux, measured by EGRET by three orders of magnitude. The latter can be approximated
as
$$F(E>E_{\gamma}) \approx 3\cdot 10^{-6} (\frac{E_0}{E_{\gamma}})^{1.1} (cm^2 \cdot s \cdot ster)^{-1},$$
where $E_0 = 451 MeV$.

The above upper bound strongly restricts ($f_4 \le 10^{-9}$ the earliest abundance
because of the consequent impossibility to reduce the primordial $U$-hadron abundance
by $U$-hadron annihilation in low density objects. In the cases A and B annihilation in
such objects should not take place, whereas annihilation within the dense objects,
being opaque for $\gamma$ radiation, can avoid this constraint due to strong suppression of outgoing
$\gamma$ flux. However, such constraint nevertheless should arise for the period of dense objects' formation.
For example, in the course of protostellar collapse hydrodynamical timescale $t_H \sim 1/\sqrt{\pi G \rho} \sim 10^{15} s /\sqrt{n}$
exceeds the annihilation timescale
$$t_{an} \sim \frac{1}{f_4 n \sv} \sim  \frac{10^{12} s}{f_4 n}
\left(\frac{1/30}{\alpha}\right)^{2}(\frac{T}{300K})^{9/10}(\frac{m}{250GeV})^{11/10}$$
at $n > 10^{14} \left(\frac{10^{-10}}{f_4}\right)^{2}\left(\frac{1/30}{\alpha}\right)^{4}\left(\frac{T}{300K}\right)^{9/5}\left(\frac{m}{250GeV}\right)^{11/5}$,
where $n$ is in $cm^{-3}$. We consider homogeneous cloud with mass $M$ has radius $R \approx \frac{10^{19}cm}{n^{1/3}} (\frac{M}{M_{\odot}})^{1/3}$,
where $ M_{\odot} = 2 \cdot 10^{33}g$ is the Solar mass. It becomes opaque for $\gamma$ radiation, when this radius
exceeds the mean free path $l_{\gamma} \sim 10^{26} cm/n$ at $n> 3 \cdot 10^{10}cm (\frac{M}{M_{\odot}})^{1/2}$.
As a result, for $f_4$ as large as in the case C, rapid annihilation takes place when the collapsing matter is
transparent for $\gamma$ radiation and the EGRET constraint can not be avoided. The cases A and B are consistent with
this constraint.

Note that at $f_4 < 5 \cdot 10^{-6} (\frac{M}{M_{\odot}})^{2/9}$, i.e. for all the considered cases
energy release from $U$-hadron annihilation does not exceed the gravitational binding energy of the collapsing body.
Therefore, $U$-hadron annihilation can not prevent the formation of dense objects but it can provide additional
energy source, e.g. at early stages of evolution of first stars. Its burning is quite fast (few years)
and its luminosity may be quite extreme, leading to a short inhibition of star formation.

According to above arguments $U$-baryon abundance
in the primary cosmic rays can be close to the primordial value $f_4$.
It gives for case B
\beq
f_4 = \frac{r_4}{r_b} \sim 3 \cdot 10^{-9}.
\label{primcr}
\eeq
If $(Uuu)$ is the lightest $U$-baryon, its
fraction in cosmic ray helium component can reach in this case the value
$$\frac{(Uuu)}{^4 He} \sim 3 \cdot 10^{-8},$$
which is accessible for future cosmic rays experiments,
such as RIM-PAMELA and AMS 02 on International Space Station.

Similar argument in the case C would give for this fraction $\sim 2 \cdot 10^{-5}$,
what may be already excluded by the existing data. However,
it should be noted that the above estimation assumes significant
contribution of particles from interstellar matter to
cosmic rays. If cosmic ray particles are dominantly
originated from the purely stellar matter, the decrease of
$U$-hadron abundance in stars would substantially reduce
the primary $U$-baryon fraction of cosmic rays.


\section{\label{gas} Galactic blowing of $U$-baryon atoms polluting our Earth}
Since the condition Eq.(\ref{lowcond}) is valid for the disc
interstellar gas, having the number density $n_g \sim 1$cm$^{-3}$ one can expect
that the $U$-hadron abundance in it can decrease relative to the
primordial value only due to enrichment of this gas by the matter,
which has passed through stars and had the
suppressed $U$-hadron abundance according to Eq.(\ref{sol2}).
Taking the factor of such decrease of the order of the
ratio of total masses of gas and stars in Galaxy $f_g \sim 10^{-2}$
and accounting for the acceleration of the interstellar gas
by Solar gravitational force, so that the infalling gas has velocity
$v_g \approx 4.2 \cdot 10^6$cm/s in vicinity of Earth's orbit,
one obtains that the flux of $U$-hadrons coming with interstellar
gas should be of the order of
\beq
I_U = \frac{f_4 f_g n_g v_g}{8 \pi} \approx 1.5 \cdot 10^{-7} \frac{f_4}{10^{-10}} (cm^{2}\cdot s \cdot ster)^{-1},
\label{interst}
\eeq
where $f_4$ is given by the Eq.(\ref{primcr1}).

Presence of primordial $U$-hadrons in the Universe should be reflected by their existence
in Earths atmosphere and ground.
However, according to  Eq.(\ref{sol2}) (see
discussion in Section \ref{matter}) primordial
terrestrial $U$-hadron content should strongly decrease due to
radiative recombination, so that the $U$-hadron
abundance in Earth is determined by the kinetic
equilibrium between the incoming $U$-hadron flux
and the rate of decrease of this abundance
by different mechanisms.

In the successive analysis we'll concentrate our attention on the case,
when the lightest $U$ baryon is doubly charged $(Uuu)$ and
the lightest $\bar U$ meson is electrically neutral $(\bar U u)$.
In this case $U$ baryons look like superheavy anomalous helium
isotopes.

Searches for anomalous helium were performed in series of
experiments based on accelerator search \cite{exp1}, spectrometry
technique \cite{exp2} and laser spectroscopy \cite{exp3}.
From the experimental point of view an anomalous helium
represents a favorable case, since it remains in the atmosphere
whereas a normal helium is severely depleted in the terrestrial
environment due to its light mass.

The best
upper limits on the anomalous helium were obtain in \cite{exp3}.
It was found by searching for a heavy helium
isotope in the Earth's atmosphere that in the mass range 5 GeV - 10000 GeV the
terrestrial abundance (the ratio of anomalous helium number to the
total number of atoms in the Earth) of anomalous helium is less
than $3 \cdot 10^{-19}$ - $2 \cdot 10^{-19}$. The search in the atmosphere is reasonable because heavy
gases are well mixed up to 80 km and because the heavy helium
does not sink due to gravity deeply in the Earth and is
homogeneously redistributed in the volume of the World Ocean at
the timescale of $10^3$yr.

The kinetic equations, describing evolution of anomalous helium and $U$-mesons
in matter have the form
\beq
\frac{dn_{U}}{dt} = j_{U} - n_{U} \cdot n_{\bar U} \cdot \sv - j_{gU}
\label{kin1}
\eeq
for  $U$-meson number density $n$ and
\beq
\frac{dn_{\bar U}}{dt} = j_{\bar U} - n_{\bar U} \cdot n_{U} \cdot \sv - j_{g \bar U}
\label{kin2}
\eeq
for number density of anomalous helium  $n_U$.
Here $j_{U}$ and $j_{\bar U}$ take into account the income of, correspondingly, $U$-baryons
and $U$-mesons to considered region, the second terms on the right-hand-side of equations
describe $U \bar U$ recombination and the terms $j_{gU}$ and $j_{g \bar U}$ determine
various mechanisms for outgoing fluxes, e.g. gravitationally driven sink of particles.
The latter effect is much stronger for $U$-mesons due to much lower mobility
of $U$-baryon atoms. However, long range Coulomb like interaction prevents them from
sinking, provided that its force exceeds the Earth's gravitational force.

In order to compare these forces let's consider the World's Ocean as a thin shell
of thickness $L \approx 4 \cdot 10^5$cm
with homogeneously distributed $y$ charge,
determined by distribution of $U$-baryon atoms with concentration $n$. The
$y$-field outside this shell according to Gauss' law is determined by
$$2 E_y S = 4 \pi e_y n S L,$$
being equal to
$$E_y = 2 \pi e_y n L.$$
In the result $y$ force, exerting on $\bar U$-mesons
$$F_y = e_y E_y,$$  exceeds gravitational force for $U$-baryon atom concentration

\beq
n > 10^{-7} \frac{m}{250\GeV}\left(\frac{30^{-1}}{\alpha_y}\right) cm^{-3}.
\label{neutr}
\eeq

Note that the mobility of $U$-baryon atoms
and $\bar U$ mesons differs by 10 order of magnitude,
what can lead to appearance of excessive $y$-charges
within the limits of (\ref{neutr}). One can expect
that such excessive charges arise due to the effective slowing down
of $U$-baryon atoms in high altitude levels
of Earth's atmosphere,
which are transparent for $\bar U$ mesons,
as well as due to the 3 order of magnitude decrease of $\bar U$ mesons
when they enter the Earth's surface.

Under the condition of neutrality, which is strongly protected by
Coulomb-like $y$-interaction, all the corresponding parameters for $\bar U$-mesons
and $U$-baryons in the Eqs.(\ref{kin1})-(\ref{kin2}) are equal, if Eq.(\ref{neutr}) is valid.
Provided that the timescale of mass exchange between the Ocean and atmosphere is much less than
the timescale of sinking, sink terms can be neglected.

The stationary solution of Eqs.(\ref{kin1})-(\ref{kin2}) gives in this case
\beq
n = \sqrt{\frac{j}{\sv}},
\label{statsol}
\eeq
where
\beq
j_{U}=j_{\bar U}=j \sim \frac{2 \pi I_U}{L} = 10^{-12} \frac{f_4}{10^{-10}} cm^{-3}s^{-1}
\label{statin}
\eeq
and $\sv$ is given by the Eq.(\ref{sigimpact}).
For $j \le 10^{-12} \frac{f_4}{10^{-10}} cm^{-3}s^{-1}$
and $\sv$ given by Eq.(\ref{sigimpact}) one obtains in water
$$ n \le \sqrt{\frac{f_4}{10^{-10}}} cm^{-3}.$$
It corresponds to terrestrial $U$-baryon abundance
$$ r \le 10^{-23}\sqrt{\frac{f_4}{10^{-10}}},$$
being below the above mentioned experimental upper limits for anomalous helium ($ r < 10^{-19}$)
even for the case C with $f_4= 2\cdot 10^{-6}$.
In air one has
$$ n \le 10^{-3}\sqrt{\frac{f_4}{10^{-10}}} cm^{-3}.$$
For example in a cubic room of 3m size there are nearly 27 thousand heavy hadrons.

Note that putting formally in the Eq.(\ref{statsol}) the value of
$\sv$ given by the Eq.(\ref{hadsigmv}) one obtains $ n \le 6
\cdot 10^2 cm^{-3}$ and $ r \le 6 \cdot 10^{-20}$, being still
below the experimental upper limits for anomalous helium
abundance. So the qualitative conclusion that recombination in
dense matter can provide the sufficient decrease of this
abundance avoiding the contradiction with the experimental
constrains could be valid even in the absence of gauge $y$-charge
and Coulomb-like $y$-field interaction for $U$-hadrons. It looks
like the hadronic recombination alone can be sufficiently
effective in such decrease. However, if we take the value of $\sv$
given by the Eq.(\ref{hadsigmv1}) one obtains $ n$ by the factor
of $\frac{m_{\rho}}{m_{\pi}} \sim 5.5$ larger and $ r \le 3.3
\cdot 10^{-19}$, what exceeds the experimental upper limits for
anomalous helium abundance. Moreover, in the absence of
$y$-attraction there is no dynamical mechanism, making the number
densities of $U$-baryons and $\bar U$-mesons equal each other
with high accuracy. So nothing seems to prevent in this case
selecting and segregating $U$-baryons from $\bar U$-mesons. Such
segregation, being highly probable due to the large difference in
the mobility of $U$-baryon atoms and $\bar U$ mesons can lead to
uncompensated excess of anomalous helium in the Earth, coming
into contradiction with the experimental constrains.

Similar result can be obtained for any planet, having atmosphere and Ocean,
in which effective mass exchange between atmosphere and Ocean takes place.
There is no such mass exchange in planets without atmosphere and Ocean
(e.g. in Moon) and $U$-hadron abundance in such planets is determined
by the interplay of effects of incoming interstellar gas, $U \bar U$ recombination
and slow sinking of $U$-hadrons to the centers of planets. (See Appendix 6)

\section{\label{correlation} Correlation between cosmic ray and large volume underground detectors' effects}
Inside large volume underground detectors (as Super Kamiokande) and in their vicinity $U$-hadron recombination
should cause specific events ("spherical" energy release with zero total momentum or "wide cone" energy release
with small total momentum), which could be clearly distinguished from the (energy release with high
total momentum within "narrow cone") effects of common atmospheric neutrino - nucleon-lepton chain
(as well as of hypothetical WIMP annihilation in Sun and Earth).

The absence of such events inside 22 kilotons of water in Super
Kamiokande (SK) detector during 5 years of its operation would
give the most severe constraint
$$n < 10^{-3} cm^{-3},$$
corresponding to $r < 10^{-26}$. For the considered type of anomalous helium such constraint would be
by 7 order of magnitude stronger, than the results of present direct searches and 3 orders above our
estimation in previous Section.

However, this constraint assumes that distilled water in SK does still contain
polluted heavy hadrons (as it may be untrue). Nevertheless even for pure water
it may not be the case for
the detector's container and its vicinity. The conservative limit follows from the condition
that the rate of $U$-hadron recombination in the body of detector does not exceed
the rate of processes, induced by atmospheric muons and neutrinos.
The  presence of clustered-like muons originated on the SK walls
would be probably observed.
The possibility to detect upgoing muon signal from $U$-hadron recombination in atmosphere is
considered in Appendix 8.

High sensitivity of large volume detectors to the effects of $U$-hadron recombination
together with the expected increase of volumes of such detectors up to $1 km^3$
offer the possibility of correlated search for cosmic ray $U$-hadrons and
for effects of their recombination.

During one year of operation a $1 km^3$ detector could be sensitive to effects of recombination
at the $U$-hadron number density $n \approx 7 \cdot 10^{-6} cm^{-3}$ and $r \approx 7 \cdot 10^{-29}$,
covering the whole possible range of these parameters, since this level of sensitivity
corresponds to the residual concentration of primordial $U$-hadrons, which can survive
inside the Earth. The income of cosmic $U$-hadrons and equilibrium between this income
and recombination should lead to increase of effect,
expected in large volume detectors.

Even, if the income of anomalous helium with interstellar gas is completely suppressed,
pollution of Earth by $U$-hadrons from primary cosmic rays is possible.
The minimal effect of pollution by $U$-hadron primary cosmic rays flux $I_U$ corresponds to the
rate of increase of $U$-hadron number density $j \sim \frac{2 \pi I_U}{R_E}$, where $R_E \approx 6 \cdot 10^8 cm$
is the Earth's radius. If incoming cosmic rays doubly charged $U$-baryons after their slowing down in matter
recombine with electrons we should take instead of $R_E$ the Ocean's thickness $L \approx 4 \cdot 10^5 cm$
that increases by 3 orders of magnitude the minimal flux and the minimal number of events, estimated below.
Equilibrium between this income rate and the
rate of recombination should lead to $N \sim jVt$ events of recombination
inside the detector with volume $V$ during its operation time $t$.

For the minimal flux of cosmic ray $U$-hadrons, accessible to AMS 02 experiment
during 3 years of its operation ($I_{min} \sim 10^{-9} I_{\alpha} \sim 4 \cdot 10^{-11} I(E)$,
where $I(E)$ is given by Eq.(\ref{eq1}), in the range of energy per nucleon $1 < E < 10 GeV$)
the minimal number of events expected in detector of volume $V$ during time $t$
is given by $N_{min} \sim \frac{2 \pi I_{min}}{R_E}Vt$. It gives about 3 events
per 10 years in SuperKamiokande ($V=2.2 \cdot 10^{10} cm^3$) and about $10^4$ events in the $1 km^3$ detector
during one year of its operation. The noise of this rate is one order and half below
the expected influence of atmospheric $\nu_{\mu}$.

The possibility of such correlation facilitates the search for
anomalous helium in cosmic rays and for the effects of $U$-hadron recombination
in the large volume detectors.

The previous discussion assumed the lifetime of $U$-quarks $\tau$ exceeding
the age of the Universe $t_U$. In the opposite case $\tau < t_U$ all the primordial
$U$ hadrons should decay to the present time and the cosmic ray interaction
may be the only source of cosmic and terrestrial $U$ hadrons (see Appendix 9).

\section{\label{accelerators} Signatures for $U$-hadrons in accelerator experiments}
The assumed value of $U$-quark mass makes the problem of its search at accelerators similar
to the case of $t$-quark. However, the strategy of such search should take into account
the principal difference from the case of unstable top quark. One should expect that
in the considered case a stable hadron should be produced.

If these hadrons are neutral and doubly charged, as we
dominantly considered above,
the last ones can be observed as the 'disagreement' between the
track curvature (3-momentum)
\beq
p = 0.3 B \cdot R \cdot Q,
\label{curv}
\eeq
where $B$ is magnetic field in $T$, $p$ is
momentum in GeV, $R$ is radius of curvature in meters, $Q$ is the
charge of $U$ hadrons in the units of elementary charge $e$, and
the energy of the track measured in the calorimeter (or energy
loss $dE/dx$). The search for such tracks at the Tevatron is
highly desirable. The expected inclusive cross section of $U$-meson production at the
Tevatron is about 0.7 pb (for $M_Q=250$ GeV) and about 0.05 pb for the
 double charged barion (in the case of U-quark); more than a half of the
events contains two non-relativistic heavy particles with the velocity
 $\beta < 0.7$.

The argument favoring $(Uuu)$ as the lightest $U$ baryon is simply
based on the mass ratio of current $u$ and $d$ quarks. On the
other hand there is an argument in favor of the fact that the
$Uud$ baryon must be lighter than the $Uuu$ one.
 Indeed, in all models the scalar-isoscalar $ud$-diquark is lighter
than the vector-isovector $uu$-diquark. One example is the model with
the effective t'Hooft instanton induced four quark interaction, which
provides a rather strong attraction in the scalar $ud$-channel and
which is absent in the vector $uu$ channel. In a more general form we
can say that the interaction in isoscalar channel (isoscalar potential)
must be stronger than that in the isovector case. Otherwise we
will obtain a negative cross section for one or another reaction since
the isovector interaction changes the sign under the replacement of
$d$-quark by the $u$-quark.

Thus it is very likely that the $Uuu$-baryon will be unstable under
the decay $(Uuu)\to (Uud)\ +\  \pi^+$. This expectation is confirmed by
the properties of the charmed baryons (where the charm quark is much
heavier than two other quarks). Indeed, the branching of the $\Sigma_c\to \Lambda_c +\pi$
 decay is about 100\% \cite{PDG}.

The uncertainty in the choice of the lightest baryon of the 4th generation
is accomplished by the uncertainty of the choice of the lightest
quark, is it $U$-quark in the analogy with the first generation,
or $D$-quark, as it follows from the analogy with second and third
generations. These ambiguities in theoretical expectations make us to
stipulate the signatures for 4th generation
in accelerator searches, which are independent of the above mentioned
uncertainties.

The expected cross section of the charge 1 heavy meson
 (like $U\bar d$) production is not too small. It is
 comparable with the $t$-quark cross section, depending
 on the mass of heavy quark. At the moment the best limits on the
 quarks of 4-th generation was given by the CDF collaboration
 using the $dE/dx$ measurements. For the quark with electric charge
 $q=2/3$ the limit is $M_U > 220$ GeV \cite{CDF}. Therefore in the present paper we
 choose for our estimates the value $M_U=250$ GeV,
 which corresponds to the production cross section of the order of 1 pb at the
Tevatron energy\footnote{The probability to recombine with
 diquark and to form the baryon is about order of magnitude
 lower.}.


Due to a very large mass (more than 150 GeV) the new heavy hadrons are
rather slow. About half of the yield is given by particles with the
velocity $\beta < 0.7$. To identify such
hadrons one may study the events with a large transverse energy (say,
using the trigger - $E_T > 30$ GeV). The signature for a new heavy
hadrons will be the 'disagreement' between the values of the full
energy $E=\sqrt{m^2+|\vec p|^2}-m$ measured in the calorimeter, the
curvature of the track (which, due to a larger momentum $|\vec
p|=E/\beta$, will be smaller than that for the light hadron where
$E\simeq |p|$) and the energy loss $dE/dx$.
For the case of heavy hadrons due to a low $\beta$
 the energy loss $\frac{dE^{QED}}{dx}$ caused by the electromagnetic interaction
 is larger than that for the ultrarelativistic light hadron, while
 in "hadron calorimeter" the energy loss caused by the strong
 interactions is smaller (than for a usual light hadron), due to a
 lower inelastic cross section for a smaller size heavy hadron,
 like $(U\bar d)$ meson.
Besides this the whole large $E_T$ will be produced by the single
isolated track and not by a usual hadronic jet, since the expected
energy of the accompanying light hadrons
 $E_T^{acc}\sim \frac{1\  GeV}{m}E_T$ is rather low.

Another possibility to identify the new stable heavy hadrons is
 to use the Cherenkov counter or the time-of-flight information.

We hope that the hadrons, which contain a heavy quark of 4-th
 generation, may be observed in the new data collected during the
 RunII at the Tevatron and then at the LHC, or the limits on the
 mass of such a heavy quarks will be improved\footnote{If the mass
of Higgs boson exceeds $2m$, its decay channel the pair of stable $Q \bar Q$ will dominate
over the $t \bar t$, $2W$, $2Z$ and invisible channel to neutrino pair
of 4th generation \cite{nuHiggs}. It may be important for the strategy of heavy Higgs searches.}.

\section{\label{discussion} Discussion}
To conclude, the existence of hidden stable or metastable quark
of 4th generation can be compatible with the severe experimental
constrains on the abundance of anomalous isotopes in Earths
atmosphere and ground and in cosmic rays, even if the lifetime of
such quark exceeds the age of the Universe. Though the primordial
abundance $r= r_4/r_b$ of hadrons, containing such quark (and
antiquark) can be hardly less than $r \sim 10^{-10}$, their
primordial content can strongly decrease in dense astrophysical
objects (in the Earth, in particular) owing to the process of
recombination, in which hadron, containing quark, and hadron,
containing antiquark, produce unstable charmonium-like
quark-antiquark state.

To make such decrease effective, the equal number density of
quark- and antiquark-containing hadrons should be preserved.
It appeals to a dynamical mechanism, preventing segregation
of quark- and antiquark- containing hadrons. Such mechanism,
simultaneously providing strict charge symmetry of quarks
and antiquarks, naturally arises, if the 4th generation
possess new strictly conserved U(1) gauge ($y$-) charge.
Coulomb-like $y$-charge long range force between quarks and
antiquarks naturally preserves equal number densities
for corresponding hadrons and dynamically supports the
condition of $y$-charge neutrality.

It was shown in the present paper that if $U$-quark is the
lightest quark of the 4th generation, and the lightest
$U$-hadrons are doubly charged $(Uuu)$-baryon and electrically
neutral $(\bar U u)$-meson, the predicted abundance of anomalous
helium in Earths atmosphere and ground as well as in cosmic rays
is below the existing experimental constrains but can be within
the reach for the experimental search in future. The whole cosmic
astrophysics and present history of these relics are puzzling and
surprising, but nearly escaping all present bounds.

Searches for anomalous isotopes in cosmic rays and at
accelerators were performed during last years.
Stable doubly charged $U$ baryons offer challenge for cosmic ray
and accelerator experimental search as well as for increase of
sensitivity in searches for anomalous helium. In particular, they seem
to be of evident interest for cosmic ray experiments, such as PAMELA
and AMS02. $+2$ charged $U$ baryons represent the low $Z/A$
anomalous helium component of cosmic rays, whereas $-2$ charged $\bar U$ baryons
look like anomalous antihelium nuclei. However, in the baryon asymmetrical
Universe the predicted amount of primordial $\bar U$ baryons
is exponentially small, whereas their secondary fluxes originated
from cosmic ray interaction with the galactic matter are
predicted at the level, few order of magnitude below
the expected sensitivity of future cosmic ray experiments.
The same is true for cosmic ray $+2$ charged $U$ baryons,
if $U$-quark lifetime is less than the age of the Universe
and primordial $U$ baryons do not survive to the present time.

The arguments for the lightest $(Uuu)$-baryon simply use the
$u$ and $d$ current quark mass ratio. These arguments are not supported
by models of quark interactions, which favor isoscalar
$(Uud)$ baryon to be the lightest among the 4th generation hadrons
(provided that $U$ quark is lighter, than $D$ quark,
what also may not be the case).
If the lightest $U$-hadrons have electric charge $+1$ and survive
to the present time, their abundance in Earth would
exceed the experimental constraint on anomalous hydrogen.
This may be rather general case for the lightest hadrons of the 4th generation.
To avoid this problem of anomalous hydrogen overproduction
the lightest quark of the 4th generation should have
the lifetime, less than the age of the Universe.

However short-living are these quarks
on the cosmological timescale
in a very wide range of lifetimes they should behave
as stable in accelerator experiments\footnote{With an intermediate scale of about
10$^{11}$ GeV (as in supersymmetry models \cite{Benakli})
the expected lifetime of $U$-(or $D$-) quark $\sim 10^6$ years is much less
than the age of the Universe but such quark is practically stable
in any collider experiments.}.

With all the uncertainties in the predicted mass spectrum of 4th generation
hadrons one can expect the cross section of their production
to be at the level of the $t$-quark cross section.
Stability of lightest hadrons and their large mass of
about 250 GeV should make them rather slow.
It offers a chance to use as their possible signature
the "disagreement" between the total energy,
measured in calorimeter, on one side, and curvature,
single-particle character of track
and energy loss, on the other side,
in the events with large transverse energy at Tevatron and LHC.

In the present work we studied effects of 4th generation
having restricted our analysis by the processes
with 4th generation quarks and antiquarks. However,
as we have mentioned in the Introduction in the considered
approach absolutely stable neutrino of 4th generation
with mass about 50 GeV also bears $y$-charge.
The selfconsistent treatment of the cosmological
evolution and astrophysical effects of $y$-charge
plasma of neutrinos, antineutrinos, quarks and antiquarks of
4th generation will be the subject of special studies.

We believe that a tiny trace of heavy hadrons as anomalous helium and stable neutral meson\footnote{Storing these charged
and neutral heavy hadrons in the matter might influence its $e/m$ properties,
leading to the appearance of apparent fractional charge effect in solid matter.
The present sensitivity for such effect in metals ranges from $10^{-22}$ to $10^{-20}$.}
may be hidden at a low level in our Universe ($\frac{n_U}{n_b} \sim 10^{-10} - 10^{-9}$)
and even at much lower level here in our terrestrial matter a density $\frac{n_U}{n_b} \sim 10^{-23}$.
There are good reasons to bound the 4th quark mass below TeV energy.
Therefore the mass window and relic density is quite narrow and well defined,
open to a final test.

\section*{Acknowledgements}

The work of K.B., M.Kh. and K.S. was performed in the framework of the State Contract
40.022.1.1.1106 and was partially supported by the RFBR grants 02-02-17490, 04-02-16073 and
grant UR.02.01.008. (Univer. of Russia). M.Kh. expresses his gratitude to Abdus Salam International
Centre for Theoretical Physics for hospitality.

\section*{\label{freezing} Appendix 1.  Freezing out of $U$-quarks}

The expansion rate of the Universe is given by the expression
\beq
H=\sqrt{ \frac{4\pi^3g_{tot}}{45}}\frac{T^2}{m_{Pl}}\approx 1.66\,g_{tot}^{1/2}\frac{T^2}{m_{Pl}},
\label{Hubble}
\eeq
which follows from the expression for critical density of the Universe
$$\rho_{crit}=\frac{3H^2}{8\pi G}=g_{tot}\frac{\pi^2}{30}T^4.$$
When it
starts to exceed the rate of quark-antiquark annihilation
\beq
R_{ann}=n_4 \sv,
\label{annih}
\eeq
in the period, corresponding to $T=T_f<m$, quarks of 4th generation freeze out,
so that their relative concentration
\beq
r_4=\frac{n_4}{s}
\eeq
does not follow the equilibrium distribution at $T<T_f$. Here the
entropy density of the Universe on the RD stage is given by
\beq
s=\frac{\rho+p}{T}=\frac{4}{3}\frac{\rho}{T}=\frac{2\pi^2g_{tot\,s}}{45}T^3
\approx 0.44 \,g_{tot\,s}T^3.
\label{entropy}
\eeq
The entropy density can be conveniently expressed in terms of the number density
$n_{\gamma}=\frac{2\zeta(3)}{\pi^2}T^3$ of thermal photons as follows
\beq
s=\frac{\pi^4g_{tot\,s}}{45\zeta(3)}n_{\gamma}\approx1.80\,g_{tot\,s}n_{\gamma}.
\eeq

Under the condition of entropy conservation in the Universe, the number density
of the frozen out particles can be simply found
for any epoch through the corresponding thermal photon number density $n_{\gamma}$. Factors $g_{tot}$
and $g_{tot\,s}$ take into account the contribution of all particle species and are defined as
$$g_{tot}=\sum_{bosons} g_{bos}\left(\frac{T_{bos}}{T}\right)^{4}
+\frac{7}{8}\sum_{fermions} g_{fer}\left(\frac{T_{fer}}{T}\right)^{4}$$
and
$$g_{tot\,s}=\sum_{bosons} g_{bos}\left(\frac{T_{bos}}{T}\right)^{3}
+\frac{7}{8}\sum_{fermions} g_{fer}\left(\frac{T_{fer}}{T}\right)^{3},$$
where $g_{bos}$ and $g_{fer}$ is the number of spin degrees of freedom for bosons and fermions, respectively.

>From the equality of the expressions Eq.(\ref{Hubble}) and Eq.(\ref{annih}) one gets
$$m/T_f\approx42+\ln(g_{tot}^{-1/2}m_pm\sv)$$
with $m_p$ being the proton mass and obtains, taking
$\sv \sim \frac{\alpha^2_{QCD}}{m^2}$ and $g_{tot}(T_f)=g_{tot\,s}(T_f)= g_f\approx80-90$,
$$T_f \approx m/30$$
and
\beq
r_4=\frac{H_f}{s_f\sv}\approx \frac{4}{g_f^{1/2}m_{Pl}T_f\sv} \approx 2.5 \cdot 10^{-14}\frac{m}{250
{\GeV}}.
\label{freez2}
\eeq
Index "f" means everywhere that the corresponding quantity is taken at $T=T_f$. Also it is worth to emphasize,
that given estimation for $r_4$ relates
to only 4th quark or 4th antiquark abundances, assumed to be equal to each
other.

The modern number density of frozen out particles
can be found from this estimation as
$$n_{mod}=r_4\cdot s_{mod}.$$
The modern entropy density $s_{mod}$ is assumed to be determined by relic photons with temperature $T$ and light neutrinos
with temperature $T_{\nu}=(4/11)^{1/3}T$. It gives
$g_{tot\,s\,mod}=43/11$ and $s_{mod}=7.04\cdot n_{\gamma\,mod}.$

Note that if $T_f > \Delta = m_D - m$, where $m_D$ is the mass of $D$-quark
(assumed to be heavier, than $U$-quark) the frozen out concentration of 4th generation quarks
represent at $T_f > T> \Delta$ a mixture of nearly equal amounts of
$U \bar U$ and $D \bar D$ pairs.

At $T < \Delta$ the equilibrium ratio
$$\frac{D}{U} \propto \exp{(-\frac{\Delta}{T})}$$
is supported by weak interaction, provided that
$\beta$-transitions $(U \rightarrow D)$ and $(D \rightarrow U)$ are
in equilibrium.
The lifetime of $D$-quarks, $\tau$, is also determined by
the rate of weak $(D \rightarrow U)$ transition, and at $t \gg \tau$
all the frozen out $D \bar D$ pairs should decay to $U \bar U$ pairs.

\section*{\label{radiative} Appendix 2. Radiative recombination}
Radiative $U \bar U$ recombination is induced by "Coulomb-like" attraction
of $U$ and $\bar U$ due to their $y$-interaction. It can be described in the analogy
to the process of free monopole-antimonopole annihilation considered in \cite{ZK}.
Potential energy of Coulomb-like interaction between
$U$ and $\bar U$ exceeds their thermal energy $T$ at the distance
$$ d_0 = \frac{\alpha}{T}.$$
In the case of $y$-interaction its running constant $\alpha = \alpha_y \sim 1/30$
\cite{Shibaev}. For $\alpha \ll 1$, on the contrary to the case of
monopoles \cite{ZK} with $g^2/4 \pi \gg 1$, the mean free path
of multiple scattering in plasma is given by
$$\lambda = (n \sigma)^{-1} \sim (T^3 \cdot \frac{\alpha^2}{T m})^{-1} \sim \frac{m}{\alpha^3 T} \cdot d_0,$$
being $\lambda \gg d_0$ for all $T < m$. So the diffusion approximation \cite{ZK}
is not valid for our case. Therefore radiative capture of free $U$ and $\bar U$ particles
should be considered.
According to \cite{ZK}, following the classical solution
of energy loss due to radiation, converting infinite motion
to finite, free $U$ and $\bar U$ particles form bound systems at the impact parameter
\beq
a \approx (T/m)^{3/10} \cdot d_0.
\label{impact}
\eeq
The rate of such binding is then given by
\beq
\sv = \pi a^2 v \approx \pi \cdot (m/T)^{9/10} \cdot (\frac{\alpha}{m})^2 \approx
\eeq
$$\approx 6 \cdot 10^{-13} (\frac{\alpha}{1/30})^{2}(\frac{300K}{T})^{9/10}(\frac{250GeV}{m})^{11/10} \frac{cm^3}{s}.$$

The successive evolution of this highly excited atom-like bound
system is determined by the loss of angular momentum owing to
y-radiation. The time scale for the fall on the center in this
bound system, resulting in $U \bar U$ recombination was estimated
according to classical formula in \cite{DFK}

\beq \tau = \frac{a^3}{64 \pi} \cdot (\frac{m}{\alpha})^2 =
\frac{\alpha}{64 \pi} \cdot (\frac{m}{T})^{21/10} \cdot
\frac{1}{m} \label{recomb} \eeq
$$\approx 4 \cdot 10^{-4} (\frac{300K}{T})^{21/10}(\frac{m}{250GeV})^{11/10} s.$$

As it is easily seen from Eq.(\ref{recomb}) this time scale of $U
\bar U$ recombination $\tau \ll m/T^2 \ll m_{Pl}/T^2$ turns to be
much less than the cosmological time at which the bound system was
formed.

Kinetic equation for U-particle abundance with the account
of radiative capture on RD stage is given by
Eq.(\ref{hadrecomb}).
Its solution at $T_0=T_f > T > T_{QCD}=T_1$, assuming in this period $g_{tot \,(s)}=const=g_f$
what allows to use transformation similar to Eq.(\ref{transform}),
is given by
\beq
r_4 = \frac{r_0}{1 + r_0 \sqrt{\frac{\pi g_f}{45}}\, m_{Pl} \int^{T_0}_{T_1}\sv dT} \approx
\frac{r_0}{1 + r_0\, \sqrt{\frac{20 g_f\pi^3}{9}}\,
\frac{\alpha^2 m_{Pl}}{m} \, \left(\frac{T_f}{m}\right)^{1/10}}
\label{radrecsolq1}
\eeq
$$ \approx 0.013\, \left(\frac{m}{T_f}\right)^{1/10} \frac{m}{\alpha^2 m_{Pl}}\approx 4\cdot 10^{-16}
\frac{m}{250\GeV}\left(\frac{30^{-1}}{\alpha}\right)^2,$$
where $r_0$ is given by Eq.(\ref{freez2}).

At $T < T_{QCD}$ the solution for the effect of radiative recombination is given by
\beq
r_4 \approx \frac{r_0}{1 + r_0\, \sqrt{\frac{20 g_{QCD}\pi^3}{9}}\,
\frac{\alpha^2 m_{Pl}}{m} \, \left(\frac{T_{QCD}}{m}\right)^{1/10}}\approx r_0
\eeq
with $r_0$ taken at $T = T_{QCD}$ equal to $r_4$ from Eqs.(\ref{hadrecsol}),(\ref{hadrecsol2}) or (\ref{hadrecsol4}).

Owing to more rapid cosmological expansion radiative capture of $U$-hadrons
in expanding matter on MD stage is less effective, than on RD stage.
So the result $r_4 \approx r_0$ holds on MD stage with even better
precision, than on RD stage. Therefore radiative capture does not change
the estimation of $U$-hadron pregalactic abundance, given by Eq.(\ref{hadrecsol}).

\section*{Appendix 3. Hadronization of $U$ quarks}
Here we show that in the baryon excessive background $U$ quarks
form $U$-baryons and $\bar U$ form $\bar U$ mesons, while the
possible abundance of $\bar U$ antibaryons and $U$ mesons is
suppressed exponentially. Indeed, even if the number density of
$\bar U$-baryons, $n_{-}$, was initially comparable with the one
of $U$-baryons, successive reactions with nucleons, such as \beq
(\bar U \bar u \bar u) + N \rightarrow (\bar U u) + 2 \pi
\label{reac} \eeq substantially reduce this number density down
to a negligible value.

The decrease of relative $\bar U$-baryon abundance,
$r_-=n_-/s$,
is described by the equation
\beq
\frac{dr_{-}}{dt} = - r_{-} \cdot n_{b} \cdot \sv.
\label{reacdec}
\eeq
Provided that total baryon number in Universe did not change significantly
since QCD phase transition, nucleon number density $n_b$ at the given time
can be defined through the relation for baryon to entropy ratio $r_b$, which reads
\beq
\frac{n_b}{s} = r_b = const=\frac{n_{b\,mod}}{s_{mod}}=\frac{1}{7.04}\frac{n_{b\,mod}}{n_{\gamma\,mod}} \approx 10^{-10}.
\label{bard}
\eeq
At modern epoch the ratio
$\eta_b = n_{b\,mod}/n_{\gamma \,mod} \approx 6\cdot 10^{-10}$.

To solve equation Eq(\ref{reacdec}) we will use the transformation
\beq
-s\cdot dt=\sqrt{\frac{\pi g_{QCD}}{45}}\,m_{Pl}\cdot dT.
\label{transform}
\eeq
This transformation of differentials is applicable if
the effective total number of degrees of freedom is
approximately a constant during the most of temperature interval of
considered period following the QCD phase transition,
$g_{tot}(T<T_{QCD})=g_{tot\,s}(T<T_{QCD})= g_{QCD}\approx const$.
Such a condition can be assumed to be fulfilled taking $g_{QCD}\approx 15$,
which is a roughly averaged value
over the most temperature interval after QCD transition. This value changes from about
17 to about 12 within 150MeV-200keV temperature interval and then virtually instantaneously
it reduces down to the modern value, which is about 4.
For our estimation we will take the cross section of such recombination as $\sigma \sim
\frac{1}{m_{\pi}^2} \cdot \frac{1}{v}$
and for rate of this reaction (\ref{hadsigmv})
$$\sv \sim \frac{1}{m_{\pi}^2} \approx 6 \cdot 10^{-16}\, {\rm \frac{cm^3}{s}}.$$
Then one obtains the exponentially strong suppression of primordial $\bar
U$-baryon abundance
\beq
r_{-} = r_0 \exp{\left ( -\int^{T_{0}}_{T_1} \frac{n_{b}}{s} \sv \sqrt{\frac{\pi g_{QCD}}{45}}\,dT \right )}
\approx r_0 \exp{\left( - 0.15\, \eta_b\, \frac{m_{Pl}}{m_{\pi}} \frac{T_0}{m_{\pi}} \right)},
\label{antisup}
\eeq
giving for $T_0 \sim T_{QCD}$ ($T_1\ll T_0$)
$$r_{-} \sim r_0 \exp{(-10^{10})}.$$

Qualitatively the result does not change if we take more conservative estimation for the rate of this reaction (\ref{hadsigmv1}).
Substituting $m_{\rho}$ instead of $m_{\pi}$ in the Eq.(\ref{antisup}) one obtains for
$T_0 \sim T_{QCD}$
$$r_{-} \sim r_0 \exp{(-4 \cdot 10^{8})}.$$

The same exponential suppression takes place for $(U \bar u)$ and $(U \bar d)$ mesons.
Therefore only $U$-baryons and $\bar U$-mesons should be considered as possible
primordial forms of $U$-hadrons.

Similar suppression of $U$-hadrons with light valent antiquarks ($\bar u$ and $\bar d$)
should take place in any dense baryonic environment.
In particular, it makes the lightest $U$ baryon and $\bar U$ meson the only possible forms
of $U$-hadrons in terrestrial, solar or selenal matter.

At $T> T_w \approx 1$MeV the rate of weak $\beta$-reactions exceeds
the rate of cosmological expansion. Owing to these reactions
and small mass difference of light $u$ and $d$ quarks
one should expect comparable amounts of $(Uuu)$, $(Uud)$ and $(Udd)$
baryons (as well as of $(\bar U u)$ and $(\bar U d)$ mesons)
in the Universe in the period $T_{QCD} > T > T_w$.

Low number density of $U$-hadrons makes negligible their
role in reactions of Standard Big Bang Nucleosynthesis.
In general one may discuss a bound deuteron-like state
 formed by the proton and $(\bar U u)$ meson in these reactions. However it looks
 unlikely that the interaction between the proton and this heavy
 'meson' is strong enough to provide such a bound state.

 Indeed, the deuteron is a weakly bound system which has a very
 small binding energy. In conventional proton-neutron potentials
 (see for example \cite{BonnB}) the attraction between two
 nucleons is described mainly by the isoscalar exchange (such as
 $\sigma$-boson exchange). Therefore, assuming the additive quark
  model, we expect the proton-$u$-quark interaction to be 3 times
 smaller than that in deuteron. This is not enough to provide the
 existence of a bound state.

  For the case of $p-\bar u$-interaction the smallness
 may be partly compensated by the $\omega$ exchange which changes
 the sign interacting with the antiquark.
  Thus the situation with the possibility to form a $(U\bar u)+p$ bound
 state is not so evident. However, as it was discussed above,
after QCD phase transition hadronic reactions such as
$$(U \bar u)+p \rightarrow (Uud)+\pi^0$$
should have resulted in exponential suppression of
$(U\bar u)$-meson number density. Moreover, even being formed,
$(U\bar u)+p$ state is unstable relative to this reaction.

\section*{Appendix 4. Hadronic recombination}

In this Appendix we evaluate in different ways the cross section for formation
 of charmonium-like $(Q\bar Q)$-meson in collisions of hadrons,
containing heavy quark $Q$ and its antiquark $\bar Q$.
We assume that such hadrons have initial kinetic energy $E \sim T$,
where $T$ is temperature. In the course of collision hadrons form "compound
hadron", in which $Q$ and $\bar Q$- quarks move with relative momentum $k_{in}$.
Heavy quarks also possess $y$-attraction with a "fine structure constant"
$\alpha_y$.

 In the formation of "compound hadron" y-attraction plays important role.
One should take into account Sakharov enhancement in the cross section,
which at $v/c \ll 1$ is reduced to $2 \pi \alpha_y c/v$.
 It reminds the classical problem of accretion, when particle falls down
 the massive body of finite size (radius $R$) in its central potential.
 The well known cross section is
 $\sigma \sim 2 \pi R(R + 2GM/v^2)$.
 In the case of y-attraction the corresponding expression reads
 $ \sigma \sim 2 \pi R(R + \alpha_y /T)$,
 with $R \sim 1/m_{\pi}$ given by the hadronic size.\\

 Thanks to the y-attraction two colliding hadrons, one of which contains
 the $Q$ quark and another one - the antiquark $\bar Q$, obtain
 the momenta $k_{in} \sim \sqrt{\alpha_y m_{\pi} m} \sim 1$ GeV
even for very low $T$.
Thus momentum of heavy quark inside
compound hadron is determined by $T_{eff} = \max{\{T, \alpha_y m_{\pi}\}}$.
For $T \le m_{\pi}$ heavy hadron with momentum $k_{in}$ crosses
the compound hadron (of the size $R \sim 1/m_{\pi}$) very slowly
(with $v \sim k_{in}/m$) at the timescale $t \sim (m/k_{in})(1/m_{\pi})$,
which is much larger than the hadronization timescale $t \sim (1/m_{\pi})$
So, on one side, relatively large momentum of heavy quarks gives
 us the possibility to neglect the effects of confinement, though,
on the other side, the process is very slow and hadronization
of light hadrons can play important role.

Since metastable hadrons of 4th generation are not discovered yet
(and may be do not exist), there is evidently absent any direct experimental
information about their interactions, and, in particular, about
the cross section of reaction of hadronic recombination.
Moreover, direct information on similar reaction for charmed
and b-quark hadrons is also absent. It makes this question
open for theoretical discussions and speculations. Such speculations,
assuming formation of $(Q\bar Q)$-meson as a slow process of
successive light hadron evaporation lead to the estimation
of the order of (\ref{hadsigmv}) or (\ref{hadsigmv1}) for
the rate of hadronic recombination.
The argument for such approach is that at low energy of colliding
hadrons $T \le m_{\pi}$, it is sufficient to emit few light hadrons
(e.g. pions) from the compound hadron to bind $Q$ and $\bar Q$ within it
and to prevent the disruption of compound hadron on $(Qqq)$ and $(\bar Q q)$
states.

The more justified alternative approach is to take into account
relatively large momentum of heavy quarks, to neglect effects of confinement
 and to consider
 the forming $(Q\bar Q)$ meson as a small size hydrogen-like system.
Such treatment can provide us realistic minimal estimation
of the recombination rate. In this approach we can
 use the well known results for recombination of hydrogen
 atom with the replacement of electron mass $m_e$ by the reduced
 mass $M=M_Q/2$ and the QED coupling $\alpha^{QED}$ by
 $\bar \alpha=C_F\alpha_s + \alpha_y \sim (4/3) \cdot 0.144 + 1/30 = 0.23$.\\
 Here $C_F=(N_c^2-1)/2N_c=4/3$ is the colour factor;
 the coupling $\alpha_s$ is evaluated at the distances $r$
  equal to the size of the ground state.
 For $M_Q=250$ GeV this gives the binding energy
 $E_i=(M \bar \alpha ^2)/(2i^2) = 3.2/i^2$ GeV \\
 ($E_1=3.2$ GeV for the ground state
 $i=1$. The corresponding momentum $k_i=\sqrt{2ME_i}=28/i$ GeV.)\\

 In this approach the 'inclusive' reaction
 \begin{equation}
     (Qqq) + (\bar Q q) \rightarrow (Q\bar Q) + ....(anything)
 \end{equation}
 is considered as the recombination of the "free" heavy quarks
 \begin{equation}
     Q + \bar Q \rightarrow (Q\bar Q)
 \end{equation}
inside the "compound hadron".
It is assumed that since the free quark recombination takes place at
rather small distances, the influence on it of confinement and hadronization of light quarks,
taking place at much larger distances with the probability equal
 to 1, can be neglected.

 Now to evaluate the cross section we can use the known result for
 the electron-proton recombination
\beq
\sigma_{rec}=\sigma_r
  =\sum_i \frac{1}{N_c} \frac{8\pi}{3^{3/2}} \bar \alpha^3 \frac{e^4}{Mv^2i^3} \frac{1}{(Mv^2/2+I_i)}
\eeq
  where
 $M$ and $v$ are the reduced mass and velocity of $Q$-quark;
 $I_i$ - ionization potential  ($I_i=I_1/i^2$)  [$I_1=E_1=3.2$ GeV].  The
 colour factor $1/N_c=1/3$ is the probability to find an appropriate
 anticolour.

 To sum approximately over 'i' we note that $\sigma_r\propto 1/i$
 for $I_i >> Mv^2/2=T_{eff}$ while at $I_i<T_{eff}$ the cross section
 $\sigma_i\propto 1/i^3$ falls down rapidly.

  So effectively the sum goes up to $i=i_{max}$
 with $i_{max}=\sqrt{3.2 \mbox{GeV}/T_{eff}}$ corresponding to
  $I_{i_{max}}=T_{eff}$.

 Thus the interpolation formula for recombination cross section reads:
\beq
 \sigma_r=(\frac{2\pi}{3^{5/2}}) \cdot \frac{\bar \alpha ^3}{T_{eff} \cdot I_1} \cdot \log{(\frac{I_1}{T_{eff}})}
                           = 1.8 \cdot (\frac{\log{(I_1/T_{eff})}}{T_{eff} \cdot I_1}) \mbox{mkbn}\cdot\mbox{GeV}^2
\eeq
and the recombination rate is given by
\beq
 \sv=(\frac{2\pi}{3^{5/2}}) \cdot \frac{\bar \alpha ^3}{T_{eff} \cdot I_1} \cdot \log{(\frac{I_1}{T_{eff}})} \cdot \frac{k_{in}}{M} \approx
0.4 \cdot (T_{eff} m^3)^{-1/2} (3 +
\eeq
$$+ \log{(T_{QCD}/T_{eff})}) \approx 0.56 \cdot 10^{-20} \log{(I_1/T_{eff})} \cdot (\frac{T_{QCD}}{T_{eff}})^{1/2}(\frac{250GeV}{m})^{3/2} \frac{cm^3}{s}.$$

  This formula is valid in the interval of not too small
$T_{eff} \gg m_{\pi}^2/2M$ (in order to neglect the confinement) and not too large
 $T_{eff} \ll I_1$ (at least $T_{eff} < I_1$).

  For $T_{eff}=4.7$ MeV (which corresponds to $k_{in}\sim 1.1$ GeV) it
 gives $\sigma_r= 0.8$ mb and $\sv = 7 \mbox{mkbn} \cdot c \cdot (250 GeV/m)^{3/2} =2.1 \cdot 10^{-19}(\frac{250 GeV}{m})^{3/2}\frac{cm^3}{s}$.

 Note that here we neglect the momentum of $Q$-quark inside the
 $(Qqq)$ or $\bar Q q$ hadrons since this momentum $k\sim 0.3\ -\  0.5$
 GeV is smaller than the incoming momentum $k_{in} > 1$ GeV,
acquired in hadron collision due to $y$-attraction.

\section*{\label{cosmicray} Appendix 5. Secondary $U$-hadrons from cosmic rays}
Astrophysical mechanisms of particle acceleration lead to
appearance of charged $U$-hadron component in cosmic rays. Neutral
$U$-hadrons can not be accelerated in this way directly, but their
$y$-charge can make them to follow accelerated electrically charged $U$-particles. Primordial
$U$-hadrons can be present in the interstellar gas, captured by Solar system.
This primary flux of cosmic $U$-hadrons, falling down the Earth should
enrich its $U$-hadron abundance.

Another source of such enrichment
could be cosmic ray interaction with Earth's atmosphere.
Let's estimate number of quarks of fourth generation that can be created in
collisions of
high energy cosmic protons with nuclei in the atmosphere of Earth. Assume for our estimation
that high energy cosmic rays dominantly contain protons. To create pair of $U$-quark
and its antiquark with mass $m$ the c.m. energy
\beq
\sqrt s  > 2m
\eeq
is necessary.

 In our
case
laboratory frame can be connected with Earth. Thus we can estimate c.m. energy
of
pair proton-nucleon as $s \approx 2m_p E_N,$
where $m_p$ is proton mass, $E_N$ is the energy of incident nucleon. Then
\beq
E_N > E_{th} \approx \frac{2 m^2}{m_p} \approx 1.3 \cdot 10^5GeV.
\label{thresE}
\eeq

Integral high energy proton flux depends on energy as \cite{Ginzburg}
\beq
 I\left( E\right) =\left\{
 \begin{array}{c}
 E^{-1.7}(cm^{2}\cdot s \cdot ster)^{-1}, \, E < E_{knee},\\
 3\cdot    10^{-10}\cdot    \left(    \frac{E}{10^6GeV}\right)^{-2.1}
 (cm^{2}\cdot s \cdot ster)^{-1},\, E > E_{knee},
 \end{array}
 \right.
 \label{eq1}
 \eeq
where $E_{knee}$ is about $3 \cdot 10^6 GeV$.

In the case of $m=250$ GeV the expected inclusive cross section of
$U$-hadron production in the proton-nucleon collision is of the
 order of 0.1 pb at $E_N$ about $4.5\cdot 10^6$ GeV and more than 1 pb at
  $E_N > 10^7$ GeV.
  To calculate the flux of $U$-hadrons in cosmic rays we have
used the known leading order parton-parton cross sections of $gg\to Q\bar Q$ and
$q\bar q  \rightarrow Q\bar Q$ convoluted with the LO MRST2001 partons
 \cite{MRST} and multiplied the result by the next-to-leading order
 K factor (of about 1.2) and by the mean atomic number of 'Air', $A_{atom}=14.5$.
The ratio of this cross section to the inelastic $p-Air$ cross section,
 parameterized in \cite{Bugaev}
  was convoluted with the proton spectrum Eq.(\ref{eq1})  (taken in differential form)
  and integrated over the incoming proton energy. This gives
\beq
 I_U=2.5 \cdot 10^{-21} (cm^{2}\cdot s \cdot ster)^{-1}.
\label{atmflux}
\eeq
About half of the flux $I_U$ comes from the region of $E_N>12.5 \cdot 10^6$ GeV
 where the cross section $\sigma_U \ge 2$ pb. The major part of the
produced heavy quarks after the hadronization forms the $U\bar q$ and $\bar Uq$ mesons.
 The probability to form a baryon in the recombination with a light diquark
is expected to be an order of magnitude smaller.

The estimated secondary flux of $U$-hadrons is by more than 12 orders of magnitude less, than the
primary $U$-baryon component of cosmic rays
if its fraction is given by Eq.(\ref{primcr}).

The production and propagation of heavy hadrons, created by cosmic rays at $E_p>1.3 \cdot 10^5 (\frac{m}{250 GeV})^2$ GeV,
will show itself as a long penetrating electromagnetic track, whose behaviour will be a combination of lepton-like
and hadron-like shower.

\section*{\label{gas} Appendix 6. $U$-baryon atom pollution in Solar system}

The $y$-charge neutrality condition holds neutral $\bar U$ mesons near
much less mobile $U$-baryon atoms in the slow sinking to the centers of planets.
Representing the form of anomalous helium atoms
$U$-baryon atoms are not chemically active and they should not form chemical
compositions with matter, which might prevent them from sinking down the center.

 Planet's gravity force on the surface is $F_g = m g$ and $U$-baryon atom of mass $m$
sinks due to its action with velocity $V = \mu F_g$, where the mobility $\mu = m n \sigma v$
is determined by matter atom number density $n$ and rate of multiple atomic collisions with matter
 $\sigma v$. It gives
\beq
V = \frac{g}{n \sigma v}.
\label{vsinksur}
\eeq
Taking for Moon $g = 1.6 \cdot 10^2 cm/s^2$; $n = 6 \cdot 10^{23}/A cm^{-3}$; $\sigma \sim 10^{-16} (m_a/m) cm^2$,
averaged atomic weight in selenal matter $A = m_a/m_p \approx 30$
and thermal velocity of atoms $v = 2 \cdot 10^4 (m/m_a)^{1/2} (T/300 K)^{1/2} cm/s$, one obtains the velocity of sinking down the
selenal surface $V \sim 10^{-8} (300 K/T)^{1/2}cm/s$.

Inside the homogeneous planet with matter density $\rho$ at the radius $R$ gravity
force is $F_g = m G \frac{4 \pi}{3} \rho R$. Then
$$V = \frac{4 \pi}{3} \frac{G m_a R}{\sigma v},$$
where $m_a$ is the mass of planet matter atom, and
the timescale of sinking down the center of planet is given by
$$t = R/V = \frac{3}{4 \pi} \frac{\sigma v}{G m_a} = 6 \cdot 10^{16} (\frac{T}{300 K})^{1/2} s$$
independent of actual matter density and radius of planet.
At internal temperature $T > 3000 K$ this timescale exceeds the age of Solar system, what means that
sinking can still continue in deep hot planet's interiors.

Taking for Moon the same flux (\ref{interst}) of $U$-hadrons falling down the Moon with interstellar gas
and taking $j \sim \frac{4 \pi I_U}{R}$ with $R \sim 10^8 cm$ being the Moon's radius one
obtains from (\ref{statsol}) the number density of $U$-hadrons in selenal matter $n \sim 0.3 cm^{-3}.$

In stars $U$-baryons should be dominantly ionized and the cross section of their collision
in plasma does not differ much from such cross section for neutral $\bar U$-mesons.
It strongly increases the velocity of $U$-hadron sinking and reduces the timescale
of sinking down to stellar center. For Sun this timescale is $ 4 \cdot  10^{9}(T/10^6 K)^{1/2} s$.
With the account for the order of magnitude increase in Eq.(\ref{interst}) of velocity of gas,
falling down the Sun, and taking for Sun $j \sim \frac{4 \pi I_U}{R_{\odot}}$ with $R_{\odot} = 7 \cdot 10^{10}cm$
one obtains from Eq.(\ref{statsol}) the number density of $U$-hadrons inside the Sun $n \sim 7(T/10^6 K)^{1/2} cm^{-3}.$
Note that the presence of doubly charged U-baryons in Solar corona could lead to anomalous He$^+$ lines,
which can be hardly observable if there is no mechanism for local increase of U-baryon abundance.

Effects of Solar wind and Solar activity prevent interstellar gas from falling down the Sun
reducing Solar abundance of U-baryons. At the distance $R$ from Sun such effects decrease as $\propto R^{-1/2}$.

Heliopause at the border of Solar system, where solar wind stops pushing out infalling gas,
may provide a shield, preventing interstellar gas penetration in Solar System. Provided that this shielding
is sufficiently effective, anomalous helium coming with interstellar gas can be stopped in heliopause,
and the effects of its pollution of Earth's matter can strongly decrease.

\section*{\label{recombination} Appendix 7. Effects of $Q$-hadron recombination in matter}
Within the effective thickness of atmosphere-Ocean layers or within the sinking depth
equilibrium between  $U$-hadron income and recombination is established.
In the result of recombination, which takes place homogeneously in all the regions
occupied by $U$ hadrons, rest energy of $U$-quarks  converts into the energy of ordinary hadrons.
The latter (dominantly $\pi$ and $K$ mesons) should in their turn give rise
to photons and neutrinos as their decay products. Total hadronic energy release
in recombination corresponds to $2m I_U$ with energy of ordinary hadrons $E < m$.
For $I_U$ given by Eq.(\ref{interst}) and $m = 250 GeV$ it is by order of magnitude
smaller than the hadronic energy released in Earth's atmosphere and ground by cosmic rays.

Moreover, the dense matter is opaque for charged pions and kaons originated from $U \bar U$
recombination. For charged pions with energy $E \sim 10 GeV$ the rate of their absorption in water
$$\Gamma_i = n \sv \approx 4 \cdot 10^8 1/s$$
is by 3 orders of magnitude higher, than their decay rate
$$\Gamma_d = \frac{m_{\pi}}{E} \frac{1}{\tau_{\pi}} \approx 5 \cdot 10^5 1/s.$$
It leads to strong absorption of charged pions and to corresponding suppression ($\sim \Gamma_d/\Gamma_i$)
of neutrino fluxes from their decay. Such fluxes would be still larger
than the effect of $U \bar U$
recombination in atmosphere, in which it is much stronger suppressed by the factor
$\propto (n_{atm}/n_w)^2 \sim 10^{-6}$, where $n_{atm}$ and $n_w$
are respectively atomic number densities in air and water.

Neutral pions can dominantly escape absorption, but deep layers of water
are opaque for gamma radiation from their decay, reducing the observable gamma
source to a surface layer of thickness about $l_{\gamma}$, mean free path of $\gamma$ in water.
All the Earth is however transparent for neutrino with energy $E \le m = 250 GeV$.

So the stationary regime of $U\bar U$ recombination in Earth should be
 accompanied by the neutrino flux and by the gamma radiation from the
 Ocean surface. The calculation of these fluxes is valid for the general case
of any choice for lightest $Q$-quark and $Q$ hadrons, what we reflect in notations
below.

  The mean multiplicities of $\nu$ and $\gamma$ produced in the
 $Q\bar Q$ recombination at rest were evaluated using the JETSET 7.4
 Monte Carlo code \cite{JETSET}. We neglect the spin-spin interaction and
 assume that just due to statistics 25\% of the $U\bar U$ pairs are in the
 pseudoscalar ($0^-$) state, and thus mainly decay onto the two gluon jets,
  while 75\% of pairs are in the $1^-$ state and decay onto the three gluon
 jets. The results are presented in Table 1 where we show the total
 multiplicities of gamma and electron, muon and $\tau$ neutrinos for the
 case of $m=250$ GeV.
  To give some impression about the energy distributions we present also the
 multiplicities of particles with the energy $E\  > $ 0.1, 1, 10 and 100 GeV.
 Besides this in Table 1 we show the energy fractions carried by the
  gamma and electron, muon and $\tau$ neutrinos as well.
   Note that the numbers for $E>100$ GeV had in our calculations a low statistical
 significance and are shown just to demonstrate the strong suppression of
 the fluxes at high energies. Moreover all the numbers corresponds to the
 $Q\bar Q$ annihilation in vacuum and do not account for the attenuation
 caused by the absorption of high energy pions in medium (Earth or Ocean).

 The last effect is crucial for the neutrinos produced via the charged
 pion decay. The lifetime of a fast pion is proportional to the Lorenz
 gamma factor $E/m_{\pi}$ and grows with energy. The density of nucleons in the Earth is about
 $n=1.5 \cdot 10^{24} cm^{-3}$ , the cross section of $\pi$ absorption
 $\sigma(\pi N) \approx 20 mb = 2\cdot10^{-26} cm^2$
 So the mean free path of pions is $l=1/(n\sigma)=30 cm$.
 On the other hand the decay length for charged pions $c \cdot \tau=780 cm \cdot (E/m_{\pi})$.
 Thus the suppression is $l/(c \tau)=(30 \cdot 140MeV/780)/E_{\pi}
 = 5.4 MeV/E_{\pi}$.
  Strictly speaking $E_{\nu} < E_{\pi}$ (so we can expect a stronger
 than $5 MeV/E_\nu$ suppression)
 On the other hand after the interaction the fast pion does not
 disappear completely; it creates few pions of a lower energy
 (due to this reason the suppression should be not so strong)
 These two factors approximately compensate each other.
 So the final estimate  $5 MeV/E_{\nu}$ looks reasonable.

Therefore in the Earth
 the yield of fast neutrinos will be additionally suppressed by the factor of
 about $5 MeV/E_{\nu}$. However the prompt neutrinos, coming mainly from the decay of charmed
 particles, are not suppressed by this effect. To demonstrate the role of
 prompt neutrinos for each sort of neutrinos in the right columns of Table 1 we present the
 multiplicities (and corresponding energy fractions) calculated in the
 limit when the charged pions, kaons and $\mu$-mesons are stable.
    For a high energies exceeding few GeV the flux of prompt neutrinos
 give the dominant contribution.

\begin{table}[h]
 \begin{tabular}[]{|c|c|c|c|c|c|c|c|}
 \hline & $\gamma$ &  \multicolumn{2}{c|} {$\nu_e+ \bar{\nu_e}$} &  \multicolumn{2}{c|}
{$\nu_{\mu}+\bar{\nu_{\mu}}$} &  \multicolumn{2}{c|} {$\nu_{\tau}+\bar{\nu_{\tau}}$} \\ \hline
$N_{total}$ &  69  & 62 & 0.095 & 124 & 0.091 & 0.02 & 0.019 \\ \hline
 $N(E>0.1 \GeV)$ & 62 & 47 & 0.094 & 97 & 0.090 & 0.02 & 0.019 \\ \hline
 $N(E>1 \GeV)$ & 28 & 14.4 & 0.080 & 31 & 0.077 & 0.017 & 0.016 \\ \hline
 $N(E>10 \GeV)$ & 2.4 & 0.61 & 0.028 & 1.5 & 0.027 & 0.0057 & 0.0057\\ \hline
 $N(E>100 \GeV)$ & 0.001 & 0.0004 & 0.00017 & 0.0004 & 0.00010 & 0.0003 & 0.000 \\ \hline
 Energy fraction & 0.27 & 0.12 & 0.0018 & 0.26 & 0.0017 & 0.0004 & 0.00035\\ \hline
 \end{tabular}
 \caption{Multiplicities of $\gamma$, electron-, muon- and $\tau$- neutrinos produced in the recombination of
$(Q \bar Q)$ pair with $m=250$ GeV. First line shows the total multiplicities while the next 4 lines correspond
to multiplicities of particles with the energies $E>0.1,\,1,\,10,\,100$ GeV (in the $(Q \bar Q)$ pair rest frame).
 The last line presents the energy fraction carried by each sort. For each sort of neutrinos numbers in the left and right columns
 correspond to all neutrinos (produced in the result of fragmentation and decays of all annihilating
 products in vacuum) and to only prompt neutrinos respectively.}
 \end{table}


In particular, the stationary regime of $Q \bar Q$
recombination in Earth should be accompanied by gamma radiation with the flux
$F(E) = N(E) I_U l_{\gamma}/L$, where energy dependence of multiplicity $N(E)$ of
$\gamma$ quanta with energy $E$ is demonstrated in table 1, and $l_{\gamma}$
is the mean free path of such $\gamma$ quanta. At $E > 100 MeV$ one obtains the
flux $F(E> 100 MeV) \sim 3 \cdot 10^{-9} \frac{f_4}{10^{-10}} (cm^2 \cdot s \cdot ster)^{-1}$,
coming from the surface layer $l_{\gamma} \sim 10^2 cm$.

One should also expect neutrino flux from $Q \bar Q$ recombination in Earth with
 spectrum given in table.
At $E_{\nu} > 1 GeV$ it corresponds to the flux $\sim 2 \cdot 10^{-8} \frac{f_4}{10^{-10}} (cm^2 \cdot s \cdot ster)^{-1}$.

\section*{\label{metastable} Appendix 8. Upgoing muon signal from $U \bar U$ annihilation in atmosphere}
Here we consider the stationary regime of equilibrium between the incoming interstellar gas pollution
and $U \bar U$ annihilation in matter and discuss the possibility to detect the upgoing muon signal from such annihilation in atmosphere.

Assuming an infalling flux, given by Eq.(\ref{interst}) $I_U \approx 1.5 \cdot 10^{-7} \frac{f_4}{10^{-10}} (cm^{2}\cdot s \cdot ster)^{-1}$,
we must observe a tiny component of annihilation in air ($\sim 10^{-3}$ as the ratio of atmospheric column height to Ocean's depth).
Effects of this annihilation will be mostly masked by a more abundant downward cosmic ray showering.

Netheretheless the presence of a tiny upgoing muons, tracking from
$U \bar U \rightarrow hadrons \rightarrow \pi^{\pm} \rightarrow \mu^{\pm}$ chains,
is the source of an "anomalous" upgoing muon flux. Its value may exceed
$$I_{\mu \uparrow} = N(E_{\mu}>1 \GeV) I_U = 4.6 \cdot 10^{-9} \frac{f_4}{10^{-10}} (cm^2 \cdot s \cdot ster)^{-1}.$$
This flux is not easy to observe on the ground, but it may be observable from mountain, airplane or baloons.

Indeed, $\mu$ trace in air exceeds 6 km its flux estimated above is already comparable with the
albedo muon flux, observed at $93^o$-$94^o$ in NEVOD-DECOR experiment
\cite{NEVOD}. Its detection may be easily tested. It may be exceeding 3-4 orders of magnitude
the upgoing muons from atmospheric $\nu_{\mu}$ ($I_{\mu} \sim (2-3)\cdot 10^{-13}(cm^2 \cdot s \cdot ster)^{-1}$).

Even $E_{\mu}>10 \GeV$ harder spectra may play an important role. The expected flux is
$I(E_{\mu}>10 \GeV) \ge 2 \cdot 10^{-10} \frac{f_4}{10^{-10}}(cm^2 \cdot s \cdot ster)^{-1}$
with negligible pollution by downward cosmic ray secondaries bent by geomagnetic fields.

These signals could be complementary or even dominant over the expected upgoing one from $\tau$ airshowers
due to UHE$\nu$, skimming the Earth's crust \cite{Fargion04,Fargion02,Jones,Fargion042}.

\section*{\label{metastable} Appendix 9. Metastable $Q$-hadrons}

Interaction of cosmic rays with the spectrum (\ref{eq1}) with
the matter in disc with averaged number density $ n \sim n_g \sim 1 cm^{-3}$
for the inclusive cross section of $U$-baryon production taken as in Appendix 5
results during the lifetime of galactic cosmic rays $T_{cr} \sim 10^7$yr
in the creation of integral cosmic ray flux of $U$ baryons
\beq
I_U \approx n \sigma_{Ub} c T_{cr} I\left( E \ge E_{th} \right)
\approx 4 \cdot 10^{-22} (cm^{2}\cdot s \cdot ster)^{-1},
\eeq
being one order of magnitude smaller than the flux (\ref{atmflux}) of $U$-hadrons produced
in the cosmic ray interaction with Earth's atmosphere.
This flux is several orders of magnitude below the
level of sensitivity expected in future cosmic ray experiments,
in particular in AMS 02.
For such low incoming flux as (\ref{atmflux}) the number density of anomalous helium stored in Earth
during its age is $n \sim 4 \cdot 10^{-12} cm^{-3}$,
which corresponds to such a low recombination rate that kinetic equilibrium is not established
and equilibrium solution (\ref{statsol}) is not valid. The anomalous helium abundance
in this case does not exceed
$r \approx 4 \cdot 10^{-35}$, what is 16 orders of magnitude
below the existing upper limits. It makes in the case $\tau < t_U$ the search for
anomalous helium at accelerators most promising.

The above estimations of primordial abundance of $U$-baryons and
of their production rate by cosmic rays are also valid for the
case, when the lightest $U$-baryon is isosinglet $(Uud)$-state
with the electric charge $+1$. However, in this case the
conclusions differ drastically from the case of double charged
baryon, since $(Uud)$-baryon represents the form of anomalous
hydrogen. Mass spectrometry was used on a variety of light
elements \cite{exp1}, \cite{exph1}, \cite{exph2}, \cite{exph3},
and achieved the lowest limit $\sim 10^{-30}$ on hydrogen
isotopic abundance at around 100 GeV. These experimental
constrains are by more than 11 orders of magnitude stronger, than
for anomalous helium. These constrains exclude the case of such
$U$-baryon with $\tau > t_U$, but in the case $\tau < t_U$ the
experimental upper limits can be satisfied, since the abundance
of anomalous hydrogen produced interaction of cosmic rays in
atmosphere and stored in Earth is in this case about $r \approx 4
\cdot 10^{-35} \frac{\tau}{t_U}$. For a very wide range of $\tau
< t_U$ this lifetime is sufficiently large to prevent the
$U$-baryon decay in detectors, making (meta-)stable $U$-baryons a
profound signature for 4th generation quarks in accelerator
searches.

If particles are sufficiently longliving to survive to the period
of galaxy formation $t_{gal}$, their annihilation can take place
and the EGRET constraint on this annihilation is valid. For $\tau
< t_{gal}$ such annihilation can not take place and the
astrophysical effects of decay should be considered. These
effects depend on the decay modes, but even for the significant
fraction of electromagnetic and hadronic cascades among the decay
products, effects of $U$-hadron decay in CMB spectrum,
electromagnetic backgrounds and light element abundance are
consistent with observational constrains (see for review
\cite{chechetkin} and \cite{book}) for the cases A and B and for
a wide range of lifetimes in the case C. In particular, for
maximal fraction of electromagnetic energy release ($\sim 0.5$)
one obtains from \cite{book} the constraint $f_4 < 5 \cdot
10^{-9} (1 +z_d)$, so that even case C is possible for the decay
redshift $z_d > 400$, corresponding to the lifetime $\tau < 5
\cdot 10^{13} s$.

Similar arguments are valid for the case, when $D$-quark is lighter than $U$-quark,
and $+1$ charged $(Duu)$ and $(\bar D u)$ are the lightest $D$-hadrons. Such hadrons
can hardly form nuclear bound states with $Z > 1$ and thus cause the problem
of anomalous hydrogen overproduction, if $D$-quark lifetime exceeds the age of the Universe.

On the other hand, though the current quark mass relation is $m_d > m_u$, it is not excluded
that the self-energy induced by the QED
 interaction can compensate the mass difference $(m_d-m_u)$, making the
electrically neutral $(\bar D d)$ the lightest $\bar D$-hadron.
If in the same time the lightest $D$-baryon is also the neutral $(Dud)$,
and the lightest $D$-hadrons are not bound with hydrogen nuclei in anomalous isotopes
of hydrogen, primordial $D$-hadrons would represent an interesting form
of strongly interacting dark matter particles.
Having only nuclear interaction with the matter, such particles decouple
from plasma at the temperatures below $T_d \sim 30$ keV, and then on MD stage,
participating the development of gravitational instability
are dominantly distributed in the halo of galaxies.

After decoupling particle velocity decreases inversely proportionally to the scale factor,
i.e. as $\propto T$. In this case the rate of radiative binding has the form
\beq
\sv = \pi a^2 v \approx \pi \cdot (m/T_d)^{9/10} \cdot (\frac{\alpha}{m})^2 (\frac{T_d}{T})^{9/5}
\label{svnoneq}
\eeq
and at $T_{RD}=T_0 < T < T_d = T_1$ solution (\ref{radrecsolq1}) of the kinetic equation (\ref{hadrecomb})
is given by

$$r_4 = \frac{r_0}{1 + r_0 \sqrt{\frac{\pi g_f}{45}}\, m_{Pl} \int^{T_0}_{T_1}\sv dT} \approx$$
\beq \approx \frac{r_0}{1 + r_0\, \frac{\sqrt{5 g_f\pi^3}}{12}\,
\frac{\alpha^2 m_{Pl}}{m} \, \left(\frac{T_c}{m}\right)^{1/10} \,
\left(\frac{T_c}{T_{RD}}\right)^{4/5}}. \label{radrecsolqnon} \eeq
Here $g_f=g_{tot\,s\,mod}=43/11$ and $r_0$ is given by
(\ref{hadrecsol}) in case A, (\ref{hadrecsol2}) in case B or
(\ref{hadrecsol4}) in the case C. In the cases A and B the
primordial abundance practically does not change, while in the
case C the $Q$-hadron abundance is reduced to the end of RD stage
by a factor of 10. The energy release, related with the reduction
of $Q$-hadron abundance in the case C, is consistent with the
constrains on the distortions of CMB spectrum, but hadronic
cascades from annihilation products can interact with primordial
$^4$He and pollute the primordial light element composition by
excessive $^6$Li, $^7$Li and $^7$Be \cite{noneq} (see review in
\cite{book}).

Indeed, following \cite{book}, hadronic jet from gluon with
energy $E_g$, originated from $Q \bar Q$ annihilation, contains
$N_{\bar p} \le \frac{1}{7} N_{\pi} \le \frac{1}{7} (E_g/1
GeV)^{1/2}$ pairs of $p$ and $\bar p$ with averaged energy
$E_{\bar p} \sim (E_g/1 GeV)^{1/2}$. The chain of nuclear
reactions with energetic $^3$He, T, D products of $p (\bar p)$
induced destruction of primordial $^4$He yields \cite{noneq}
$$\frac{\Delta n_{^6Li}}{n_b} = (E_{\bar p}/1 GeV)^{2/3}N_{\bar p} \sim 30 \cdot 2.5 \cdot 10^{-6} f_4,$$
which should not exceed the observed $^6$Li abundance
$\frac{n_{^6Li}}{n_b} \sim  10^{-10}$. It restricts the relative
amount $f_4$ of annihilated $Q \bar Q$ as $f_4 < 1.3 \cdot
10^{-6}$, what excludes the possibility of $Q \bar Q$
annihilation on late RD stage in the case C (for which $f_4 = 2
\cdot 10^{-6}$). Detailed analysis of nucleosynthesis by hadronic
cascade, similar to the considered recently in \cite{Kohri} would
only strengthen this restriction.

On the MD stage at $T_{RD} > T$ before the period of galaxy formation the rate of radiative binding
in nearly homogeneously expanding matter retains the form (\ref{svnoneq}), but the transformation (\ref{transform})
has now the form
\beq
-s\cdot dt=\sqrt{\frac{\pi g_{QCD}}{45}}\,m_{Pl}\,\sqrt{\frac{T}{T_{RD}}}\cdot dT.
\label{transform1}
\eeq
It leads to the solution of the kinetic equation (\ref{hadrecomb}) at $T_{RD}=T_0 > T > T_{gal}=T_1$ given by

$$r_4 = \frac{r_0}{1 + r_0 \sqrt{\frac{\pi g_f}{45}}\, m_{Pl} \int^{T_0}_{T_1}\sv\,\sqrt{\frac{T}{T_{RD}}} dT} \approx$$
\beq \approx \frac{r_0}{1 + r_0\, \frac{\sqrt{20 g_f\pi^3}}{9}\,
\frac{\alpha^2 m_{Pl}}{m} \, \left(\frac{T_c}{m}\right)^{1/10} \,
\left(\frac{T_c}{T_{RD}}\right)^{4/5} \,
\left(\frac{T_{RD}}{T_1}\right)^{3/10}}. \label{radrecsolqnonMD}
\eeq So on MD stage to the period of galaxy formation at $T_{gal}
= T_{CMB} \cdot (1 + z_{gal}) \sim 20 T_{CMB}$ for $T_{RD} =
T_{CMB} \cdot (1 + z_{RD}) \sim 2 \cdot 10^4 T_{CMB}$ (where
$T_{CMB}=2.7 K$ is the modern temperature of CMB) $Q$-hadron
abundance in the case C should decrease by additional factor of
20, making its pregalactic abundance about $r_4 \sim 10^{-8}$.
Such $Q$-hadron annihilation on the MD stage not only leads to
energy release close to the constrains on CMB spectrum
distortions, but also results in the $\gamma$ background
incompatible with the EGRET data. Thus in the case C the
considered neutral $Q$-hadrons should be unstable with the
lifetime, less than $10^{15}$s. Note that for the cases A and B,
which are not excluded by the above arguments, owing to high
peculiar velocity effects of $Q$-hadron annihilation in Galaxy
are strongly reduced.

Finally, an interesting case of the lightest $(Ddd)$-baryon with electric $-1$ charge can be mentioned.
At the temperatures about tens keV such baryons can bind with protons and helium nuclei
in atom-like systems. The ($(Ddd)+$p)-system with the size of $\sim 3 \cdot 10^{-12}$cm
can hardly have atomic cross section for its interaction with matter. As for the ($(Ddd)+$He)-system
its recombination with electrons makes it to follow atomic matter in the period of galaxy
formation. Specific properties of these systems may provide them to be elusive for
the existing methods of anomalous hydrogen searches.
However, this case is not supported by the mass ratios of current quarks or by the arguments of quark model.

Following the additive quark model arguments one can not expect the deuterium-like state
to be formed by $(\bar D d)$ with protons. The potential of $(\bar D d)$-nucleon
 interaction is expected to be about 3 times smaller than that
 for $p-n$ interaction.

For the case of $(Dud)$-nucleon interaction the potential is about 2/3 of the $p-n$
 potential and the factor (2/3) is not enough
 to make definite conclusion. The problem is that the effective
 reduced mass $M_{eff}$ is in this case twice larger than that for
 $p-n$ system, while the true parameter is not the potential $V(r)$
  but the magnitude $V(r) \cdot M_{eff}$. So, following the arguments
of the Additive Quark Model neutral $(Dud)n$ or $+1$ charged "nuclear" states $(Dud)p$
 are not excluded and most probably should exist.
 Their existence would exclude $D$-quark with lifetime exceeding the age of the Universe,
even if in the course Standard BBN reactions they dominantly transform into $Z > 1$
states, since such transformation can hardly reduce the abundance of $Z = +1$ states
by 15 orders of magnitude, what is necessary to escape the problem of anomalous hydrogen
overproduction. So $D$-quarks with lifetime less than the age of the Universe,
being hardly eligible
to searches for anomalous isotopes in the matter and in cosmic rays,
can become a challenge for accelerator search.

\end{document}